# *In situ* diffraction study of catalytic hydrogenation of $VO_2$: Stable phases and origins of metallicity

Yaroslav Filinchuk[1]*, Nikolay A. Tumanov[1], Voraksmy Ban[1], Heng Ji[2], Jiang Wei[3], Michael W. Swift[4], Andriy H. Nevidomskyy[2]*, and Douglas Natelson[2,5]*

[1]Institute of Condensed Matter and Nanosciences, Universite Catholique de Louvain, Louvain-la-Neuve, Belgium

[2]Department of Physics and Astronomy, Rice University MS 61, 6100 Main St., Houston, TX 77005, USA

[3]Department of Physics and Engineering Physics, Tulane University, New Orleans, LA 70118, USA

[4]Department of Physics, University of California, Santa Barbara, Santa Barbara, CA 93106, USA

[5]Department of Electrical and Computer Engineering, Rice University, MS 366, 6100 Main St., Houston, TX 77005, USA

**ABSTRACT:** Controlling electronic population through chemical doping is one way to tip the balance between competing phases in materials with strong electronic correlations. Vanadium dioxide exhibits a first-order phase transition at around 338 K between a high temperature, tetragonal, metallic state (T) and a low temperature, monoclinic, insulating state (M1), driven by electron-electron and electron-lattice interactions. Intercalation of $VO_2$ with atomic hydrogen has been demonstrated, with evidence that this doping suppresses the transition. However, the detailed effects of intercalated H on the crystal and electronic structure of the resulting hydride have not been previously reported. Here we present synchrotron and neutron diffraction studies of this material system, mapping out the structural phase diagram as a function of temperature and hydrogen content. In addition to the original T and M1 phases, we find two orthorhombic phases, O1 and O2, which are stabilized at higher hydrogen content. We present density functional calculations that confirm the metallicity of these states and discuss the physical basis by which hydrogen stabilizes conducting phases, in the context of the metal-insulator transition.

## INTRODUCTION

Vanadium dioxide is a strongly correlated material with a first-order transition at 338 K between a tetragonal metallic phase at high temperatures and a monoclinic semiconducting phase at low temperatures, in the absence of strain[1]. As in $TiO_2$ and other transition metal oxides, atomic hydrogen may be intercalated into $VO_2$, where it acts as a dopant. Methods for preparing hydrogenated $VO_2$ include electrochemical insertion[2], conversion from water-containing paramontroseite[3], and catalytic spillover[2a, 4]. Beyond mere doping of the semiconducting state, investigators have reported that hydrogen doping of $VO_2$ strongly affects the metal-insulator transition, stabilizing a metallic phase[2b, 3-4] over a broad temperature range. Understanding how hydrogen alters the metal-insulator transition (MIT) in $VO_2$ can shed light on the underlying physics of the MIT and is particularly relevant given recent reports of attempts to modulate the MIT in electrochemical environments[5].

In this paper we present *in situ* structural studies of the intercalation and deintercalation of atomic hydrogen in vanadium dioxide powder, aided by minute quantities of catalytically active Pd nanoparticles. By thermally cycling hydrogenated $VO_2$ powders in both inert and hydrogen atmospheres, x-ray diffraction patterns obtained at a synchrotron allow us to determine the phase diagram for the solid solution of hydrogen in $VO_2$. While both the monoclinic and tetragonal structures of $VO_2$ are able to dissolve some hydrogen, at larger H concentrations the tetragonal phase is converted into one of two orthorhombic, metallic phases (referred to as O1 and O2) that are stable over a broad temperature range. The structural information of these two orthorhombic phases, previously unknown, was obtained for the first time with the help of synchrotron-based x-ray studies and neutron diffraction, the latter allowing the characterization of hydrogen ordering in the O2 phase. The high resolution structural studies and phase diagram obtained in the present work extend the candidate phase diagram for the $H_xVO_2$ system (beyond the $x$<<1 regime examined in studies of material obtained by conversion of paramontroseite[3]) to much higher hydrogen concentrations.

## EXPERIMENTAL SECTION

**Sample preparation and synthesis of O1 phase by low-pressure hydrogenation.** Pure vanadium dioxide powders were prepared starting with commercially supplied $VO_2$ material (Alfa Aesar). As-supplied material was further purified by annealing under mildly reducing conditions (5 mbar of Ar) at 900°C for 30 minutes. The resulting crystallites, tens of microns in diameter, were examined by optical microscopy and confirmed via their change in optical properties to undergo the MIT near the conventional transition temperature. These crystallites were mechanically ground using mortar and pestle to a finer powder with typical grain size ~ 10 microns. After mixing with a small amount of pure Si as an internal standard, a representative sample of this powder was examined at room temperature via conventional powder x-ray diffraction (Rigaku DMAX Ultima II), and the resulting diffraction pattern was an excellent match with established lattice parameters for M1-phase $VO_2$, as determined by JADE 9.4 analysis software.

The $VO_2$ powder was then used to prepare $VO_2$/Pd material. The powder was thoroughly mixed with a solution of Pd nanoparticles suspended in water (Sciventions Pd nanoparticle suspension 1.5 mg/mL), such that the mass fraction of Pd relative to $VO_2$ was 1%. The mixture was then heat-dried and vacuum-desiccated. Previous studies of hydrogenation of $VO_2$ via catalytic spillover[4, 6] had found that diffusion of atomic hydrogen proceeds most rapidly along the tetragonal *c* (monoclinic *a+c*) direction. The sample preparation was intended to increase the likelihood that most if not all $VO_2$ grains were in contact with multiple sources of catalytic spillover, while still maintaining a mass fraction for structural characterization dominated by the $VO_2$. Powder XRD measurements of the $VO_2$/Pd material indeed showed patterns identical to those produced with pure $VO_2$ powders, with any Pd signal falling below the detection threshold.

Some of the $VO_2$/Pd material was then exposed to molecular hydrogen by annealing under flowing forming gas (20% $H_2$ and 80% $N_2$, total 1 bar) at 190 °C for 10 hours in a tube furnace before being cooled to room temperature. Additional material was prepared by sealing this material under 90 psi of forming gas in stainless steel tubing and baking at 190 °C for one week in an oven. For the rest of this manuscript, we refer to this material as $H_xVO_2$/Pd. The resulting material exhibited powder XRD data consistent with an orthorhombic *Pnnm* phase previously assigned as isostructural to β-$HCrO_2$.[2a] The subsequent studies reported below involving *in situ* characterization of the structure used these three powders ($VO_2$, $VO_2$/Pd, and $H_xVO_2$/Pd) as their starting points.

***In situ* synchrotron X-ray powder diffraction experiments** were carried out at two synchrotrons and using a laboratory system composed of the MAR345 area detector, rotating anode Mo Kα radiation and XENOCS focusing mirror. Samples of the powders described above were placed into 0.5 mm thin-walled glass capillaries and fixed on a goniometric head either with wax, permitting partial release of any pressure build-up due to possible hydrogen desorption, or, for the experiments under controlled hydrogen pressure, using a special sample holder[7]. The latter was connected to a gas dosing system capable of providing vacuum and hydrogen pressures up to 200 bar.

Experiments with the area (2D) detector revealed very spotty Debye rings for the non-hydrogenated $VO_2$/Pd sample, and less spotty patterns due to smaller average domain size for the hydrogenated $H_xVO_2$/Pd. The variable-temperature data were collected on the $H_xVO_2$/Pd sample using a one-dimensional (1D) strip detector Mythen II, installed at the Materials Science Beamline at SLS, PSI (Villigen, Switzerland) and 0.826900 Å wavelength radiation. The sample was heated using STOE oven from the ambient temperature to 468 K at 4 K per minute rate, held at 468 K for 20 minutes, and cooled at the same rate back to the ambient temperature. During this temperature cycling, 100 powder patterns were collected extending up to 120° in 2θ, with very high angular and structural resolutions. However, the accuracy of the diffraction intensities was low, as manifested by random variations of Bragg peaks' intensities within the series of consecutive patterns. Despite spinning the capillary during the measurements to attempt directional averaging, the relatively poor particle statistics combined with the "narrow view" of the 1D detector resulted in relatively poor powder average.

A second round of experiments was made at Swiss-Norwegian Beam Lines (SNBL) of the European Synchrotron Radiation Facility (ESRF, Grenoble, France) using a pixel area detector PILATUS 2M. An integration of diffraction rings considerably improved the powder average (the accuracy of the diffraction intensities), leaving an access to relatively high angular and structural resolutions thanks to the variable sample-to-detector distance. Data were collected on the oscillated capillaries with 0.822570 Å wavelength radiation. Temperature was controlled using Oxford Cryostream 700+ or Cyberstar oven. The 2D images were azimuthally integrated using Fit2D program and data on $LaB_6$ standard[8]. A sample was typically heated under a given hydrogen pressure from the ambient temperature to 468 K at a rate fixed to 4 to 10 K/minute, the temperature was held for 20 minutes, then the sample was cooled at the same rate. We denote the specific experimental protocols for the *in situ* diffraction studies as follows:

a. $VO_2$/Pd sample was measured under 25 bar of $H_2$. The sample was heated from slightly above the room temperature to 468 K, the temperature was held at 468K for 20 minutes and then lowered to 300K. In total, 165 powder patterns were collected.

b. Another series of data was taken on the same sample under 25 bar of $H_2$ within 483-623 K, at the end holding the temperature at 623 K. 51 patterns were collected.

c. $VO_2$/Pd sample was measured under 100 bar of $H_2$ at temperatures from 308 to 468K, held at 468K for 20 minutes and cooled to 300 K. In total, 180 powder patterns were collected.

d. $H_xVO_2$/Pd sample was measured under 25 bar of $H_2$ at temperatures from 319 to 468K, held at 468K for 20 minutes and cooled to the room temperature. In total, 205 powder patterns were collected.

e. High angle data were collected for pristine $VO_2$, $VO_2$/Pd and on $H_xVO_2$/Pd at room temperature, as well as the finely ground $VO_2$/Pd sample (4 patterns).

f. $H_xVO_2$/Pd sample was measured under 1 bar of air in a closed capillary at temperatures from 316 to 468K, held at 468K for 20 minutes and cooled to 80 K. In total, 185 powder patterns were collected.

g. $H_xVO_2$/Pd was measured under 1 bar of $H_2$ at temperatures from 308 to 523K, held at 523 K for 20 minutes and cooled to 306 K. In total, 153 powder patterns were collected.

Owing to the azimuthal integration of the 2D images, the accuracy of the integrated diffraction intensities proved to be very high (i.e. data showed "good powder average"), allowing for structure determination and high quality Rietveld refinements. The main difficulty of the experiment was linked with the non-equilibrium nature of some samples: analysis of the data and our manometric deuteration experiments (see below) revealed relatively poor kinetics of hydrogenation of the hydrogen-free $VO_2$/Pd even at high pressures. In contrast, hydrogen release and uptake in the hydrogen-loaded material (hydrogenated beforehand in an autoclave) is much faster and can be observed on the time scale of minutes, allowing us to draw conclusions regarding on phase equilibria boundaries.

**Synthesis of O2 phase by autoclave hydrogenation**. To investigate the high hydrogen content limit, we performed high-pressure autoclave hydrogenation (here we define it as an *experimental protocol (h)*): ~100 mg of $VO_2$/Pd sample was placed inside stainless steel autoclave, tightly connected to a gas dosing system. In order to avoid a possible reaction between the sample and steel, the former was placed inside a glass tube, open on one end. The autoclave was placed inside an oven and the system was evacuated to ~$10^{-2}$ mbar for two hours. Then 15 bar of hydrogen was loaded and the autoclave was heated to 468 K over ~1 hour, and kept at this temperature for two days. After this period, the power was switched off and the sample was let in the oven to cool down to room temperature under 15 bar of hydrogen pressure. The autoclave was opened a day later, and the sample was quickly (within 1 hour) characterized by X-ray powder diffraction. The data revealed a difference between the samples prepared at 1 bar and 15 bar, calling for the detailed study of the latter. We note that these samples, sealed in a glass bottle under air, are stable at room temperature.

**Single-crystal and powder synchrotron X-ray diffraction study of the O2 phase.** The resulting 15 bar $H_xVO_2$/Pd sample was examined via synchrotron studies at room temperature, both in powder form and in detail using two single crystals isolated from the powder. The PILATUS 2M area detector and 0.682525 Å radiation wavelength at SNBL, ESRF were used. The crystals contained a few domains, all with identical structure; however, the good resolution of the diffraction spots permitted reliable intensity integration for the largest domains. Complete datasets were obtained up to high angles, with $R_{int}$ and $R_\sigma$ close to 1%, using CrysAlisPro software[9]. The structure was solved by direct methods (SHELXS) and refined by least-squares method in SHELXL[10].

**Manometric autoclave deuteration and neutron powder diffraction on the O2 phase.** Finally, to prepare samples for neutron powder diffraction, we performed manometric high-pressure autoclave deuteration (here we define it as an *experimental protocol (i)*). A large (1.83 gram) $VO_2D_x$/Pd sample was prepared by exposing $VO_2$/Pd sample at 468 K to 5 bar of deuterium pressure for 30 hours and then to 17 bar $D_2$ for another 4 days. Pressure measurements were done continuously using 300 bar Keller pressure gauge with 30 mbar precision. The deuterium uptake took nearly 3 full days, and then the pressure stabilized. The autoclave was slowly cooled to the room temperature. X-ray powder diffraction showed that the sample was very similar to the one obtained by high-pressure hydrogenation.

The deuterated sample was enclosed into a vanadium cylinder and measured using the HRPT instrument at the Swiss Spallation Source, SINQ, PSI, Villigen. The measurement was performed at room temperature, using 1.49385 Å wavelength neutrons. All powder diffraction data were analyzed using Fullprof Suite[11].

## RESULTS AND DISCUSSION

We first turn to the results of the *in situ* x-ray diffraction experiments. As we pointed out above, hydrogenation of the hydrogen-free $VO_2$/Pd sample is slow even at higher pressures. Heating such a sample in a 1 bar hydrogen environment and holding for 20 minutes at 486 K (a protocol determined by experimental constraints) is not sufficient to complete the hydrogenation, leading to inhomogeneous material (experimental protocols (a) and (b)). Therefore our analysis is largely focused on the structural evolution under different temperatures and hydrogen pressures of the pre-hydrogenated $H_xVO_2$ material.

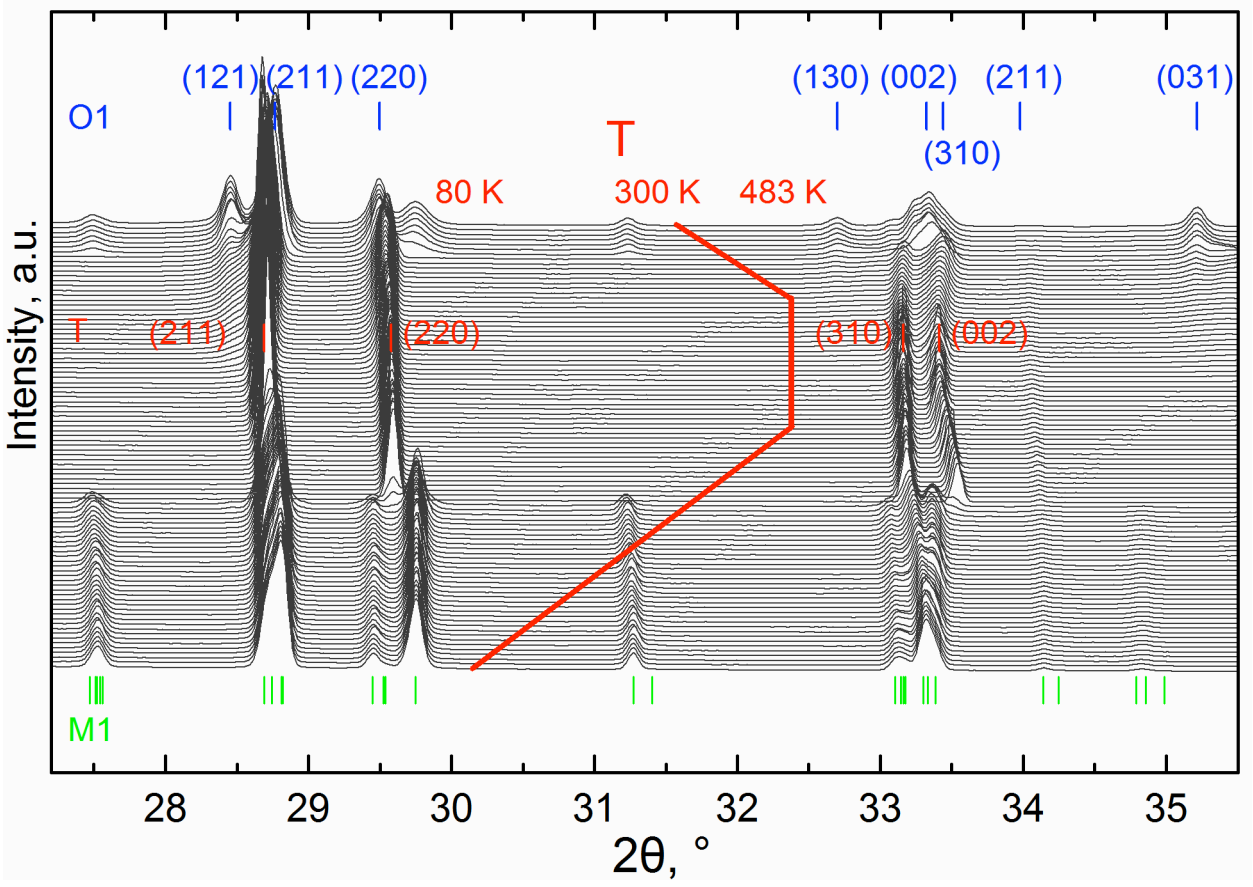

**Figure 1**. Waterfall plot of a fragment of diffraction pattern for $H_xVO_2$/Pd sample in 1 bar air, over the temperature cycle shown in red. Initially (top-most trace) the sample is a mixture of the orthorhombic O1 phase and the monoclinic M1 phase. As the sample passes through ~332 K on warming, the M1 phase transitions to the T phase. The pattern evolves with time during the high temperature part

of the cycle, as hydrogen leaves the sample. Upon cooling through ~327 K, the entire sample volume converts to the M1 phase as the temperature is further reduced. λ = 0.82257 Å.

Diffraction patterns were acquired as the sample temperature was ramped. These patterns were analyzed iteratively, and the identified phases were refined by Rietveld method until a consistent model resulted in a good fit to all data. Sequential refinement implemented in Fullprof[11] allowed for serial refinement against a number of datasets, provided the same crystalline phases were present in the given temperature range. This tedious analysis yields a complete picture on structure of crystalline phases, their evolution with temperature and on the phase transitions, and was used by us to study other hydrogen-rich systems[12].

We present the data first in the form of waterfall plots. **Figure 1** shows the diffraction pattern for the $H_xVO_2$/Pd sample under 1 bar of air, between 27° and 35° in 2θ, as the temperature is ramped (*experimental protocol (f)*). The structural phase transitions are readily apparent. In Fig. 1, the beginning of the experiment corresponds to the pattern on the top, collected at room temperature. It shows the monoclinic *P*2$_1$/*c* $H_xVO_2$ phase (M1) in a mixture with an orthorhombic *Pnnm* $H_xVO_2$ phase (O1). The M1 component transforms to the tetragonal $H_xVO_2$ (T) at ~332 K; the reverse transition from the tetragonal $VO_2$ to the monoclinic $VO_2$ occurs at 327 K on cooling. The last pattern is made only of the monoclinic phase, containing apparently very small (or no) amount of hydrogen (as determined by comparison with diffraction from the unhydrogenated, pure $VO_2$ material). We did not observe crystalline palladium or its hydrides in any of the diffraction patterns. The Rietveld refinement profile and detailed unit cell and bond length parameters for the monoclinic phase at 80 K are shown in the Supporting Information (Fig. S1 and following text). At the highest temperature of 468 K, a pure tetragonal *P*4$_2$/*mnm* $H_xVO_2$ phase is obtained. Again, the full Rietveld refinement and structural parameters are available in the Supporting Information (Fig. S2 and the following text).

The structure determination of the *Pnnm* orthorhombic phase was made on the initially prepared two-phase $H_xVO_2$ sample obtained by the hydrogenation at low pressures (the O1 phase is present in the initial traces in Fig. 1). The known structure of the monoclinic phase was fixed, refining only the cell parameters, and the remaining peaks were indexed in the orthorhombic cell. The O1 structure is a derivative of the tetragonal cell formed by an orthorhombic distortion, and corresponds to the *Pnnm* phase originally put forward by Chippindale *et al.*[2a]. In that original identification, the structure of this phase was not determined from x-ray diffraction data, though a model was suggested. The Rietveld fit of the two-phase mixture (Supporting Information, Fig. S3) is highly satisfactory, confirming this model. The refined mass fractions of the two phases are close to 1:1 in the initial trace of Fig. 1. The slightly larger formula volume of the orthorhombic cell suggests this phase is hydrogenated. Hydrogen position cannot be determined from this diffraction data alone; in the original work[2a] hydrogen positions were inferred from an inelastic neutron scattering study.

**Figure 2** shows the structures of these three phases, omitting the hydrogen positions. The structures are visualized looking just slightly misaligned from the monoclinic *a*+*c* direction (the tetragonal and orthorhombic *c*-axis). The dimerization of the V atoms that leads to the unit cell doubling of the M1 phase is readily apparent. We note that in all of the x-ray data collected, there is no evidence of detectable oxygen deficiency (put forward as a concern[5e] regarding ionic liquid gating experiments that show low temperature stabilization of a metallic T-like phase[5c]).

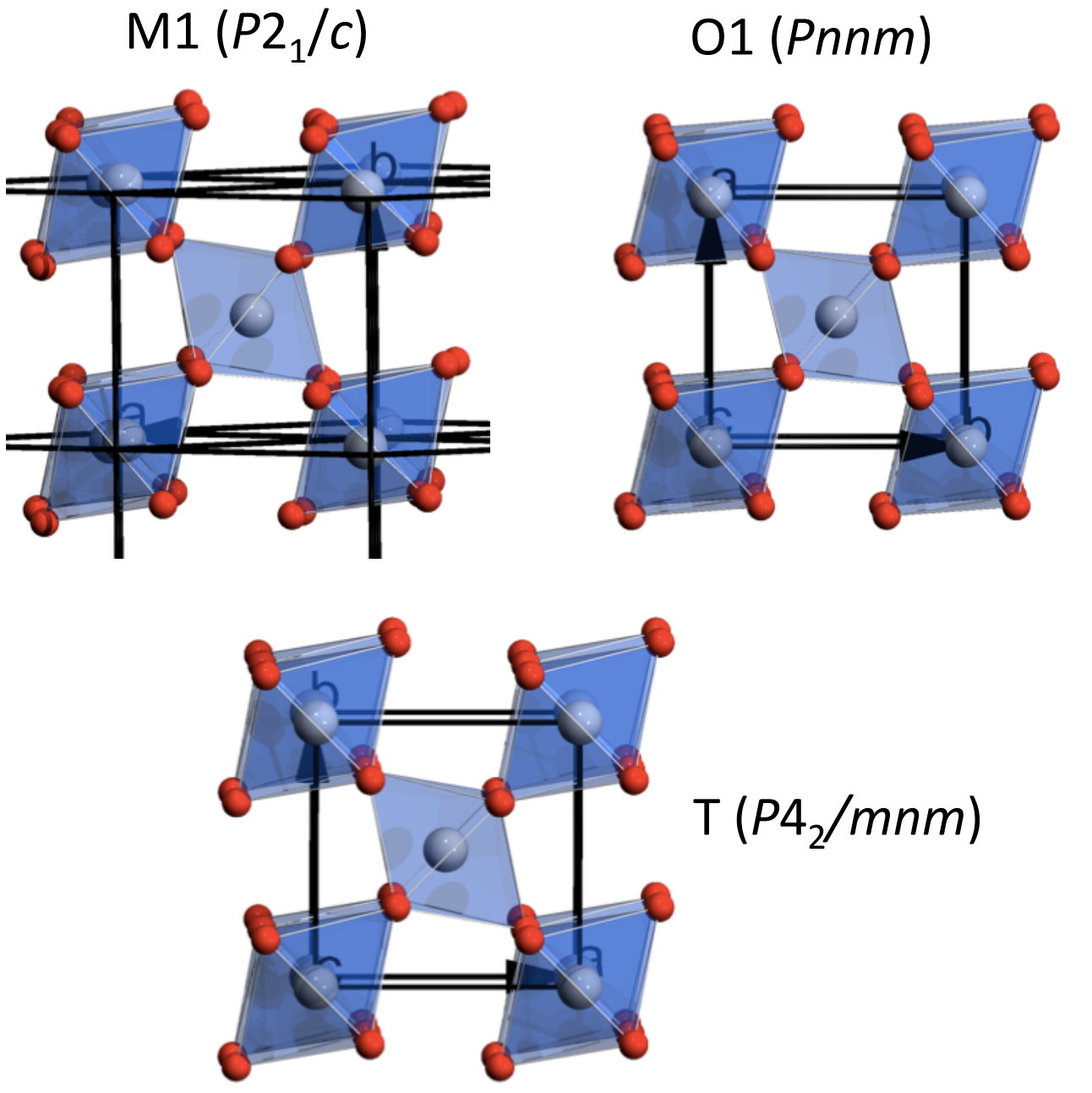

**Figure 2**. Structures of the three phases observed in the experiment of Fig. 1. Oxygen atoms are in red, vanadium atoms are colored grey. Structures are visualized from a vantage point slightly misaligned with the conventional *a*+*c* direction for monoclinic phase (tetragonal and orthorhombic *c* axis). The orthorhombicity (inequity of the *a* and *b* lattice parameters) of the O1 phase is difficult to resolve by eye at this scale. As prepared, and in the initial trace of Fig. 1, the $H_xVO_2$ sample is in a mixture (roughly 1:1) of the M1 and O1 phases. After holding at 468 K, the entire sample is converted in to the T phase, and upon cooling below 327 K the sample transforms entirely into the M1 phase.

Having determined the base structures of the relevant phases, it was possible to analyse all of the data in that context, yielding variation of cell parameters and phase fractions as a function of temperature and hydrogen pressure. **Figure 3** traces the evolution of the $H_xVO_2$ formula unit volume with temperature for Fig. 1 (*experimental protocol (f)*). Volumes generally increase with increasing temperature due to thermal expansion. The O1 fraction is gradually converted into T phase at high temperatures (as inferred from the diffraction analysis), while the volume of the T phase decreases with time while temperature is fixed at 468 K. We ascribe both of these effects to the loss of hydrogen from the O1 phase and then the T phase. Upon cooling, the volume of its $H_xVO_2$ formula unit decreases. Below 327 K, the hydrogen-depleted tetragonal phase transforms back to the monoclinic phase. There is some evidence of limited hydrogen reabsorption in the residual T phase before the last of the T phase transforms into M1, since the sample was in a sealed capillary.

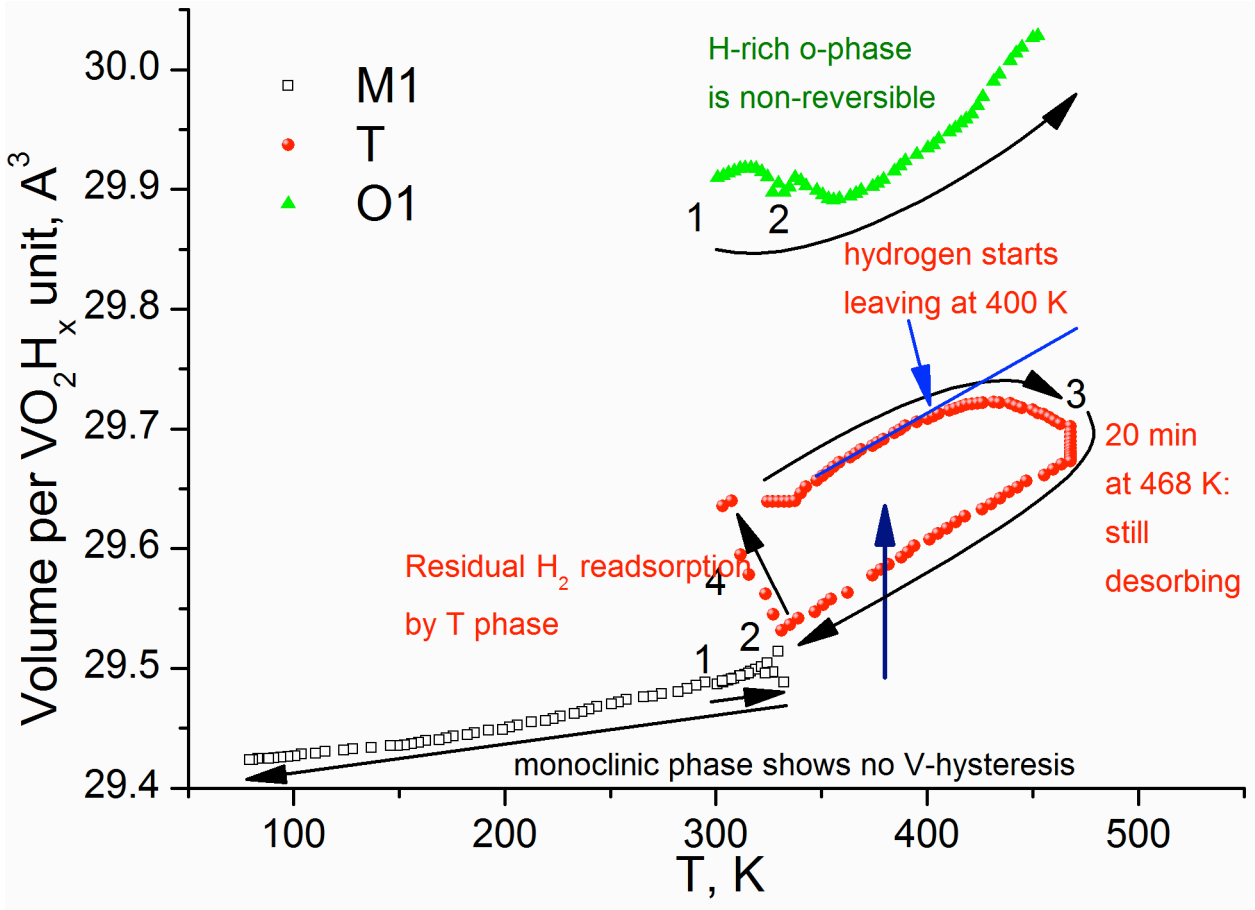

**Figure 3**. Evolution of formula unit volume throughout the heating cycle shown in Fig. 1. Initial composition (***1***) is a 50%-50% mixture of M1 (grey) and O1 (green) phase at 300 K. Upon warming (***2***), the M1 phase transforms in to the T phase (red) at 332 K. At higher temperatures, the O1 phase gradually transforms into the T phase (***3***). Upon cooling, the T phase material transforms back into the M1 phase. The conversion of O1 into T and the decrease in T formula unit volume at the high temperatures are consistent with the loss of hydrogen from the material. The related phase fractions are shown in Fig. S4.

Comparison of experimental protocols (f) and (d) provides further evidence that the hysteresis in the formula unit volume changes of the T phase are a consequence of hydrogen uptake/release. **Figure 4** traces the formula volume of the tetragonal phase for protocol (f) (no hydrogen back pressure = red squares), and for protocol (d) (one under 25 bar $H_2$ pressure = black squares). In the hydrogen-rich environment, the volume of the T phase increases irreversibly upon cycling to the high temperature, consistent with enhanced hydrogen uptake. It is clear that the tetragonal phase has a large concentration range of hydrogen solubility. At the higher hydrogen pressures the tetragonal phase absorbs more hydrogen during the heating/cooling cycle to 468/291 K, expanding its formula unit volume by 1.3% (relative to the monoclinic phase at room temperature). Upon cooling, this hydrogen-enriched tetragonal phase remains stable regarding its transformation into the monoclinic phase, down to at least 291 K. This is consistent with the observations of $H_xVO_2$ prepared by thermal decomposition of paramontroseite[3], where Wu *et al*. found a room temperature stable T phase for $x$> 0.003.

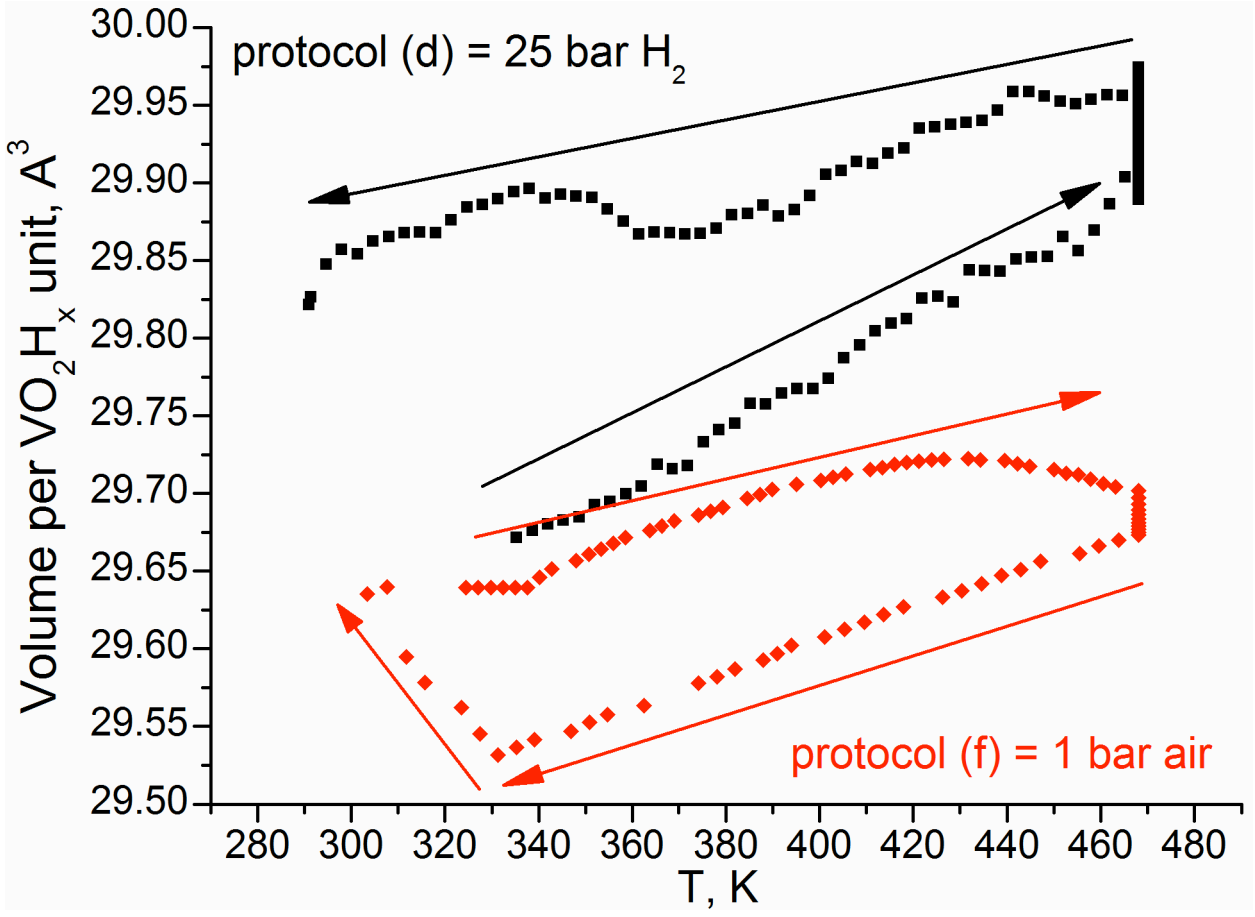

**Figure 4**. Evolution of T phase formula unit volume with temperature for different gas environment. Red squares (lower trace) correspond to protocol (f) and the data from Figs. 1 and 3. In a closed capillary with 1 bar air, the T phase expands due to thermal expansion, and contracts as a function of time due to hydrogen loss, even while held at constant temperature of 468 K. Upon cooling, volume contracts as expected, with an upturn as T phase is converted to M1 phase, likely due to hydrogen re-uptake. In contrast, black squares (upper trace) correspond to protocol (d). Under 25 bars $H_2$ gas in a closed capillary, the T phase increases its volume upon warming due to a combination of thermal expansion and hydrogen uptake.

In addition to the volume expansion of the T phase, protocol (d) reveals that hydrogenation at 25 bar induces a conversion of the tetragonal phase into a more hydrogen-rich orthorhombic *Pnnm* phase, O1. This is the same O1 phase that was already present in the as-prepared $H_xVO_2$ material (as in the initial trace in Fig. 1). Initially under 25 bar $H_2$, during the 20 minutes at 468 K, roughly 40 weight percent of the T-phase $H_xVO_2$ is transformed from the *P*$4_2$*/mnm* to the O1 *Pnnm* phase. Longer exposure to hydrogen at this temperature allows obtaining pure *Pnnm* phase. The latter is stable down to at least 80 K, and displays ~1.5% volume expansion compared to the monoclinic phase at room temperature. The *Pnnm* phase has a narrow solubility range: on cooling under high $H_2$ pressure the volume returns close to the original values. According to Chippindale *et al.*,[2a] higher hydrogen content in the orthorhombic phase leads to a more pronounced orthorhombic deformation, reaching 2.8% in our experiments.

Even higher hydrogen content leads to another orthorhombic phase. Long-term (~3 days) exposure of $VO_2$/Pd to 15 bar of hydrogen pressure at 468 K (protocol (h)) results in the formation of a more hydrogen-rich $H_xVO_2$/Pd phase, characterized by the increased unit cell volume. (We note that the kinetics of hydrogenation are comparatively poor when starting from the pure $VO_2$/Pd phase, with up to 4 days required for the complete hydrogenation (deuteration) of the pristine $VO_2$/Pd. As seen in Figs. 3 and 4, hydrogen uptake in the already-hydrogenated material is much faster and can be observed on the timescale of minutes.) Laboratory X-ray diffraction on the autoclave-hydrogenated sample shows a powder pattern similar to the one of the *Pnnm* phase, but with significantly different refined cell parameters: *a* = 4.4795(12), *b* = 4.7372(11), *c* = 2.8944(5) Å. The *b* parameter is notably increased, as compared to *a* = 4.5061(1), *b* = 4.6300(1), *c* = 2.86721(8) Å for the O1 *Pnnm* phase, thus augmenting the orthorhombic deformation. This sample has been studied in detail by diffraction on single crystals. Indexing of the reciprocal lattice unequivocally shows doubling of the orthorhombic *Pnnm* cell in all three directions. The resulting cell is *F*-centred. The systematic extinctions give only one option for the space group - the noncentrosymmetric *Fdd*2. The centrosymmetric *Fddd* is excluded by many *hk*0 reflections with *h* + *k* not equal to 4*n* having significant intensity.

Examination of the powder diffraction data revealed that the *Fdd*2 phase (O2) is a distinct new phase forming at higher hydrogen concentrations, and it is not identical (though closely related) to the *Pnnm* phase we observed at lower hydrogen concentrations. The group subgroup relation between O1 and O2 has two steps, each of the index 2: *Pnnm* (*a,b,c*) → *Pnn*2 (*a,b,c*) → *Fdd*2 (2*a*,2*b*,2*c*). The discrimination between *Pnnm* and its non-centrosymmetric subgroup *Pmm*2 for the O1 phase is much more difficult, as no systematic absence of reflections is involved. Therefore from our data we cannot claim with certainty the space group *Pnnm* for the O1 phase. It is possible that the O1 structure is pseudo-centrosymmetric with space group symmetry *Pnn*2, involving small ordered atomic displacements not detectable by powder diffraction. In that case the O1 → O2 transition occurs to the maximal isomorphic subgroup, and the transition can be of the second order.

The structure of the O2 phase was solved by direct methods and refined by the least-square method. For the best of the two crystals $R_1$ = 4.8% with 32 parameters, 316 observations and one constraint fixing the origin of the polar space group. The asymmetric unit contains two independent V atoms located on the 2-axis and two O atoms in general sites. No oxygen deficiency was detected by refining respective occupancies. The difference Fourier map revealed only one peak (4th highest) among the first 15, that was not too close to the heavy atoms (the effect of series termination). This peak was tentatively attributed to hydrogen, its coordinates were fixed, and the occupancy and the ADP (anisotropic displacement parameters) factors were refined. The excellent O1-H1 distance of 1.14 Å, the reasonable orientation of the O1-H1 bond and an appropriate refined atomic displacement factor all indicate that the peak reveals a true H position. No density was found next to the second oxygen atom, O2. The large uncertainly on the H1 occupancy does not allow quantitative determination of the hydrogen content, *x*, from the x-ray diffraction data alone.

The final results (*a* and *b* directions are swapped in order to get a conventional cell) from this analysis (further details in Supporting Information, including the Rietveld analysis) indicate cell parameters *a* = 9.4285(19) *b* = 8.9309(13), *c* = 5.7652(6) Å. In this cell there are two inequivalent V...V distances, V1-V2, of 2.783(2) and 2.982(2) Å. They are significantly different from the average 2.867 Å in the O1 structure at room temperature. The shorter of the two is longer than 2.6139(8) Å in the monoclinic phase at room temperature, which suggests that Peierls dimerization is less pronounced in the hydrogenated samples. Remarkably, the shorter V...V distance forms within $V_2O_2$ diamonds formed by H-free oxygen atoms, while the longer V...V distances appear in $V_2(OH)_2$. Thus, the separation of V...V distances fully supports localization of the H-atom only on one of the two oxygen sites (site B in Figure 10). In other words, it would appear that the hydrogen positions in the O2 phase are spatially ordered. The graphical presentation of the hydrogen ordering on the transition from O1 to O2 phase is shown in **Figure 5**.

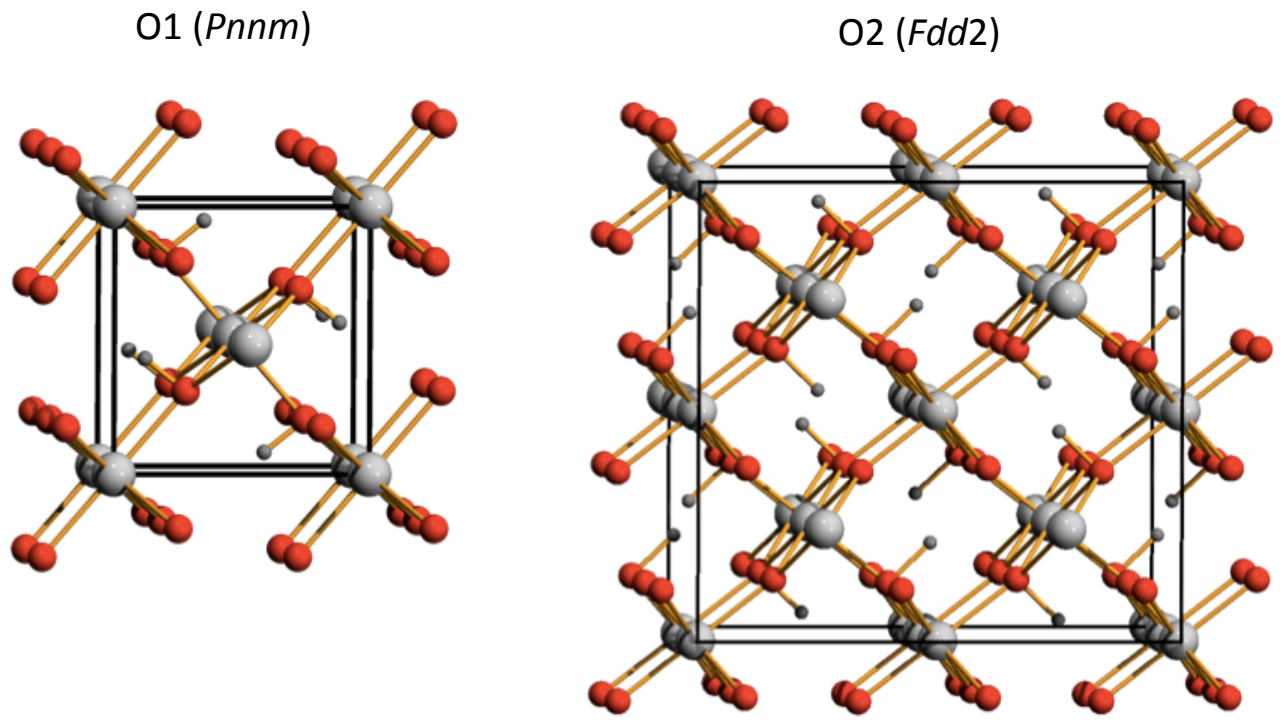

**Figure 5**. Structures for the O1 and O2 phases, as inferred from synchrotron diffraction measurements, looking nearly down the respective *c*-axes. Oxygen atoms are red, vanadium atoms are large grey, and hydrogen atoms are small grey spheres. In the O1 phase, the hydrogen atom positions are generally not ordered, and are only shown here in an ordered pattern for illustrative purposes. In the O2 phase, the hydrogen atoms are spatially ordered as shown, as is validated further in neutron powder diffraction studies of deuterated O2 material.

The strongest superstructure peaks are for *hkl* = 115, 117 and 113. Their intensities reach slightly more than 1% of the strongest peaks, 220 in *Fdd*2 or 110 in *Pnnm*. This relative peak intensity should be easily detectable by powder diffraction, so the synchrotron powder diffraction study was made on the same sample. The powder sample was measured at two sample-to-detector distances: 144 and 444 mm, corresponding to higher structural and angular resolution, respectively. Analysis of the high resolution data revealed that the sample is slightly inhomogeneous. The peaks' splitting and asymmetry was very well modelled by two identical *Fdd*2 phases, with slightly different cell parameters. The fit is stable with all cell parameters refined simultaneously. Then the model was transferred to the high-angle data and the structure was refined by Rietveld method. The final Rietveld refinement profile is shown in Supporting Information (Fig. S5, along with the zoom on the weak peaks in the low-angle part of the diagram), as are the final coordinates and quality of fit from the synchrotron powder data. The superstructure peaks are clearly visi-ble and well modelled. There is a small amount of an unidentified impurity phase(s), which does not complicate the analysis.

To reliably determine the hydrogen atom position and hydrogen content, we performed neutron powder diffraction on the corresponding deuteride DxVO2 (deuterium is used instead of hydrogen to avoid the large inelastic neutron scattering of the latter). The *Fdd*2 structure and hydrogen ordering in the O2 phase were fully confirmed, giving the refined composition $D_{0.460(8)}VO_2$, the Rietveld plot is in Fig. S6. The refined atomic positions show O1-D1 distance of 1.021(8) Å, typical for a hydroxyl group. The O-D group is capping three V atoms by oxygen, with O-D vector perpendicular to the V triangle. The nearest D-D distance exceeds 3 Å. Notably, refinement of the anisotropic atomic displacement parameters for the deuterium atom leads to the reasonable 50% probability ellipsoid (see **Figure 6**), as well as an improvement of the fit, reducing $\chi^2$ from 11.1 to 10.7. This anisotropy suggests a hydrogen hopping path from O1 to O2 atom, shown by the blue dashed bond (D1...O2 2.553 Å). An analogous site-to-site path[13] is thought to apply to diffusion of hydrogen within tetragonal $TiO_2$.

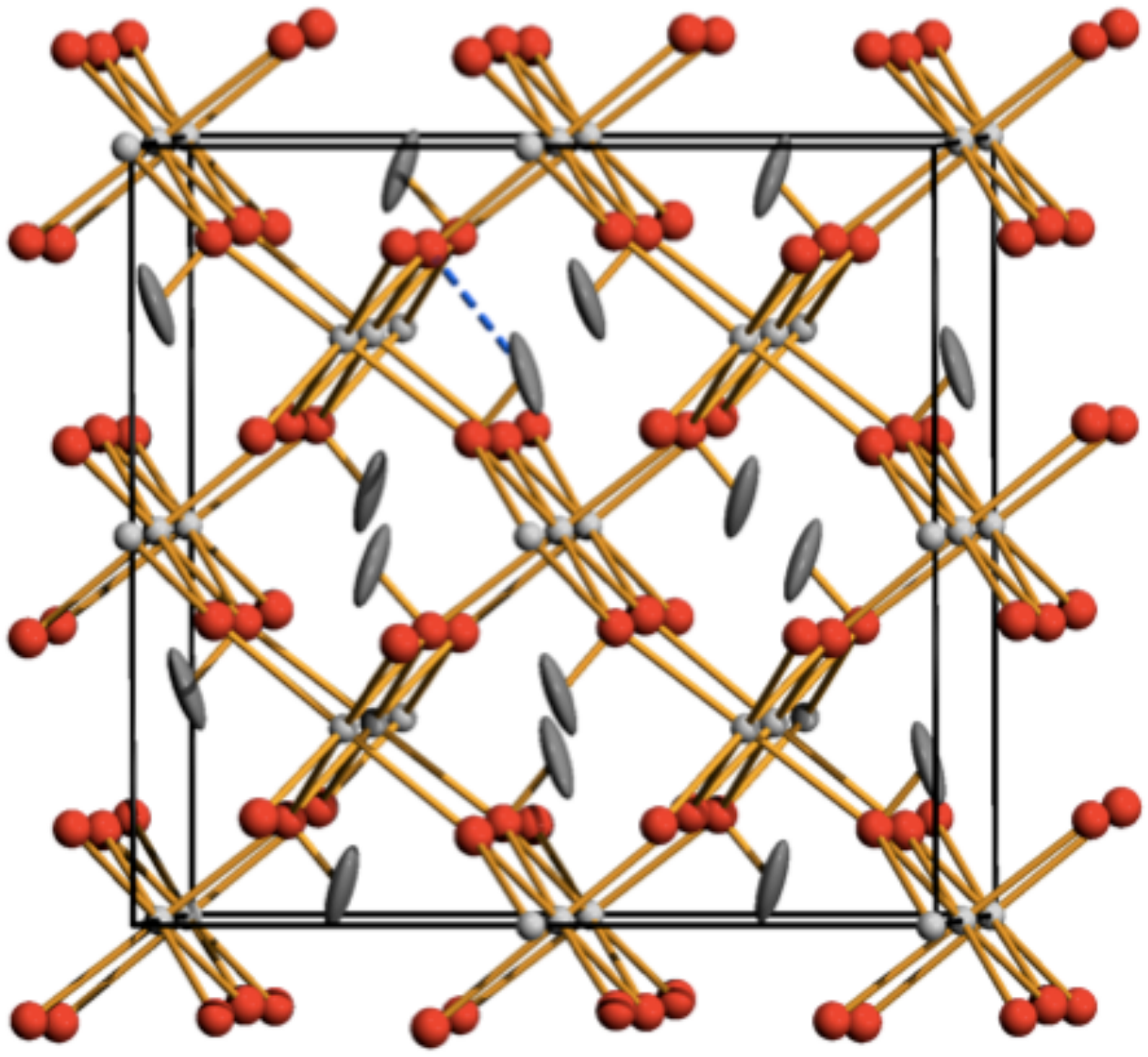

**Figure 6**. Structure of the O2 phase as determined by neutron powder diffraction. Grey 50% probability ellipsoids show the locations of the (ordered) deuterium atoms. Again the structure is drawn looking nearly down the *c*-axis. The anisotropy in the probabilistic D positions suggests the dashed blue path as a likely trajectory for site-to-site diffusion of hydrogen within the lattice.

The direct determination of the deuterium position in the O2 phase and the established group-subgroup relation with the structure of the O1 phase allows us to confirm the hydrogen position in the O1 phase proposed in Chippindale *et al.*[2a]. In the O1 *Pnnm* phase, only one independent oxygen site is present; thus the hydrogen atom is inherently disordered in the average O1 structure over the channel of oxygens running along the *c*-axis, forming a mutually exclusive H...H contact of ~1.2 Å. Supporting Information shows the final atomic coordinates and the neutron Rietveld pattern.

These data allow additional cross-checks of the relationship between orthorhombic distortion and hydrogen content. The analysis of the *b*/*a* ratios, given in the table and the graph below, summarizes data from Chippindale et al.[2a] and selected results from this work. A linear correlation between the orthorhombic distortion *b*/*a* and the hydrogen content can be established from the literature data. It suggests for the H-loaded O2 phase the composition $H_{0.6}VO_2$, consistent with the refined deuterium content in the deuterated O2 case, $D_{0.460(8)}VO_2$. Our data show that the O2 phase is characterized by 2.8-3.7% increase of the formula unit volume compared to the H-free monoclinic phase, and by 5-6% orthorhombic deformation of the parent tetragonal structure. The O1-phase demonstrates small distortion, of the order of 1.5-3%.

**Table 1. Summary of lattice parameters for various phases from prior results and this work.**

| Sample | a | b | c | β | b/a |
|---|---|---|---|---|---|
| **From Chippindale et al.[2a]** | | | | | |
| 0 | 4.554 | 4.554 | 2.847 | | 1 |
| 0.08, conventional monoclinic* | 5.366 | 4.533 | 5.372 | 118.29 | |
| 0.16 | 4.525 | 4.609 | 2.863 | | 1.0186 |
| 0.3 | 4.505 | 4.633 | 2.869 | | 1.0284 |
| 0.37 | 4.499 | 4.712 | 2.833 | | 1.0473 |
| 1 | 4.446 | 4.862 | 2.962 | | 1.0936 |
| **H-loaded O2 phase, this work** | | | | | |
| Fresh sample, subcell | 4.4795 | 4.73718 | 2.89436 | | 1.0575 |
| ESRF sc1 Fdd2 supercell | 8.9309 | 9.4285 | 5.7652 | | 1.0557 |
| ESRF sc2 Fdd2 supercell | 8.9362 | 9.4752 | 5.7791 | | 1.0603 |
| ESRF powd phase1 Fdd2 | 8.96562 | 9.44361 | 5.78186 | | 1.0533 |
| ESRF powd phase2 Fdd2 | 8.9695 | 9.49971 | 5.7925 | | 1.0591 |
| **Deuteride by neutrons at room T, this work** | | | | | |
| $VO_2D_{0.460(8)}$ | 8.9705 | 9.4610 | 5.7813 | | 1.0547 |
| **Various other phases, this work** | | | | | |
| M1, H-free, RT | 5.352 | 4.52115 | 5.38048 | 115.208 | |
| M1, hydrogenated, 301 K | 5.35716 | 4.52418 | 5.3805 | 115.249 | |
| O1, 301 K | 4.50607 | 4.63002 | 2.86725 | | 1.0275 |
| T, lower H-content, 331 K | 4.55269 | | 2.84958 | | |
| T, higher H-cont (25 bar $H_2$), 331 K | 4.57426 | | 2.857 | | |

*The published (non-conventional) monoclinic cell as $a_P$ = 5.759, $b_P$=4.533, $c_P$ = 5.372 Å and $\beta_P$ = 122.49°. The conventional cell axes are related to the published as $-a_P$-$c_P$, $-b_P$ and $c_P$.

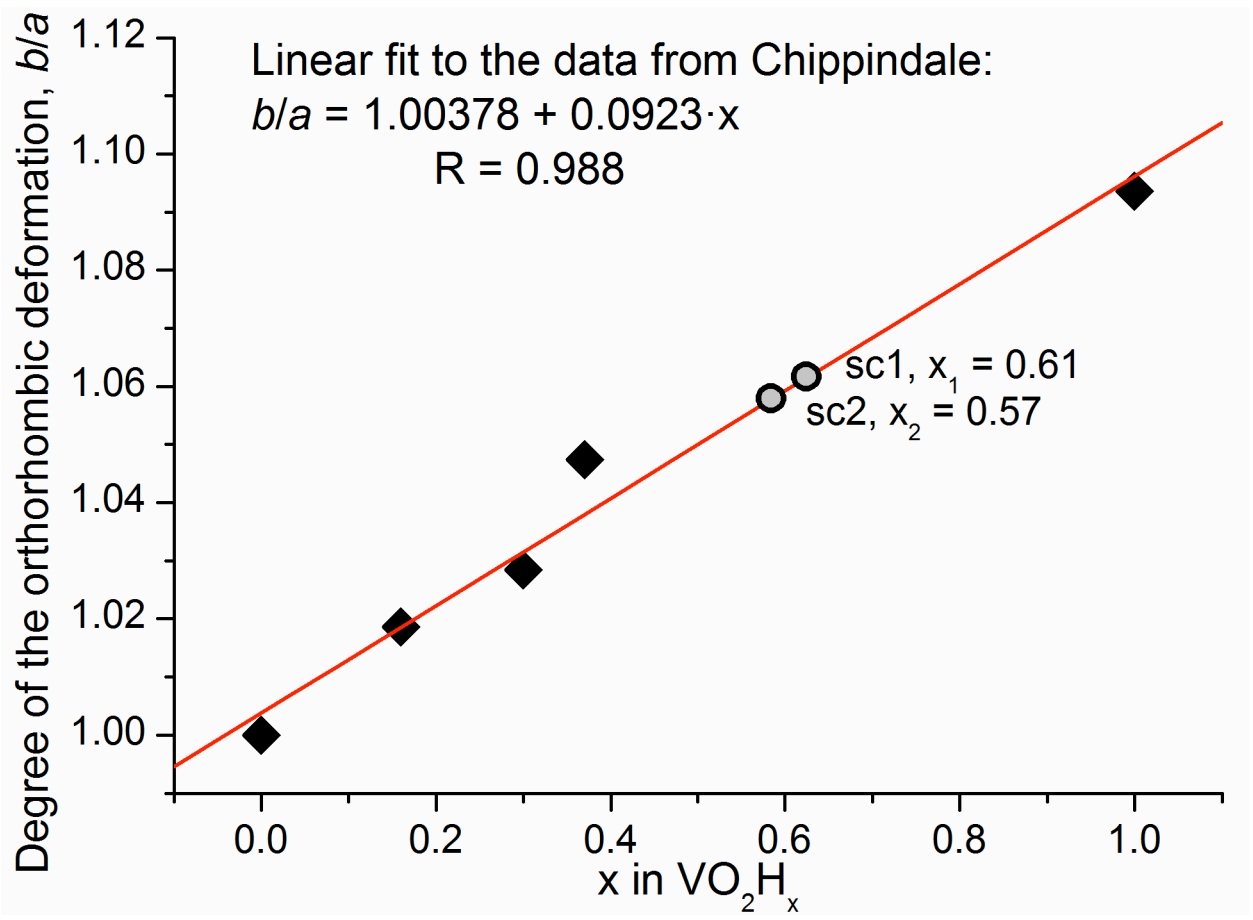

**Figure 7**. Orthorhombic anisotropy (*b*/*a* ratio) as a function of hydrogen content. Data from Chippindale *et al.*[2a] (squares) and inferred from the ESRF powder diffraction experiments (circles).

We can combine our results to construct a phase diagram (**Figure 8**) for the $VO_2$-hydrogen system. The tetragonal phase is stable at elevated temperatures only. Its stability range increases with hydrogen concentration. However, at lower temperatures, a two-phase mixture of the H-poor monoclinic and relatively H-rich O1 phase is observed.

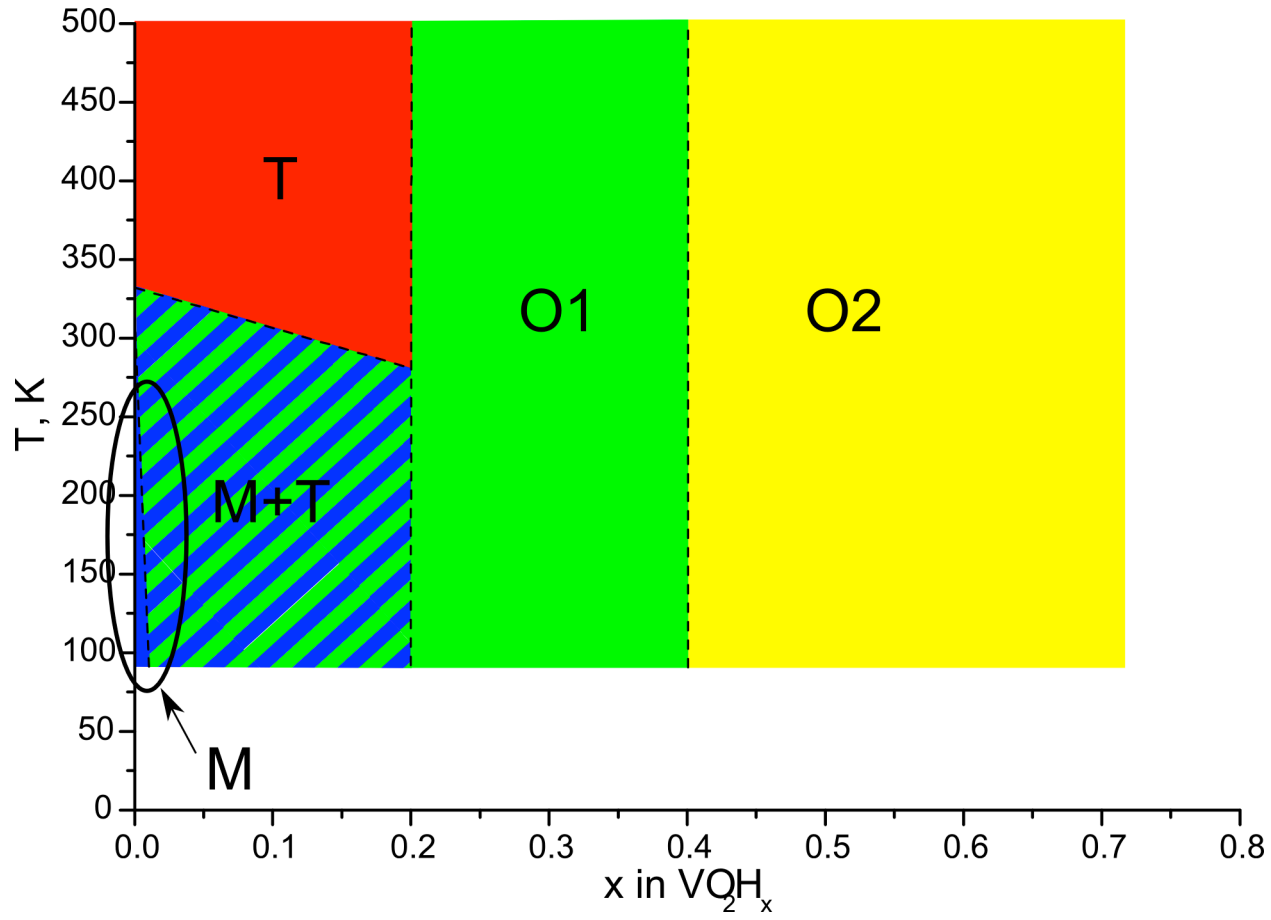

**Figure 8**. Phase diagram of the hydrogen-$VO_2$ system, deduced from the data presented in this work. The boundaries of the O1 phase are conjectural.

**THEORETICAL MODELING AND DISCUSSION.** To gain further insights into both the O1 and O2 phases, we have performed *ab initio* calculations based on the two determined structures. *Ab initio* calculations were done with Density Functional Theory (DFT) as implemented in the CASTEP[14] package. The Generalized Gradient Approximation (GGA) was used with the exchange-correlation functional of Perdew, Burke, and Ernzerhof (PBE)[15]. Ultrasoft pseudopotentials were used for all atomic species, with a plane-wave basis cutoff of 380 eV for high precision. A regular mesh of *k*-points was used to sample the Brillouin zone, 4×4×6 for the O1 phase and 2×2×3 for the O2 phase. As mentioned above, the O1 structure reported by x-ray scattering is disordered, meaning that hydrogen sites are randomly distributed throughout the lattice. Because neutron scattering data are averaged over many unit cells, this effectively results in only one inequivalent O atom connected to a hydrogen atom, on average. However to obtain the correct stoichiometry $V(OH_{x/2})_2$ of the O1 phase, the hydrogen sites are assigned partial occupancies *x/2*. In the theoretical calculations, however, we deal with a single unit cell, and such partial occupancies are impossible. Therefore, we used an ordered version of the O1 structure with $x$=0.25, with 1 in every 8 oxygens in $VO_2$ bonded to hydrogen. This of course results in non-equivalent O positions, and it was therefore essential to perform an *ab initio* structural optimization prior to calculating the electronic properties.

We note that traditional DFT methods have a notoriously difficult time reproducing both the relative energies of candidate structural ground states (M1 vs. T) and the lack of magnetic ordering in $VO_2$.[16] However, given the experimentally determined structures, DFT methods have been very successful in giving insights into the underlying physics (*e.g.*, overall electronic structure, trends with doping[4]).

The theoretical results comparing the energies of the O1 and O2 structures are shown in **Figure 9c.** The binding energy of the O1 structure is lower for small hydrogen concentrations $x \leq 0.25$, however at higher concentrations, the O2 structure is energetically favoured, which agrees well with the experimental findings presented earlier.

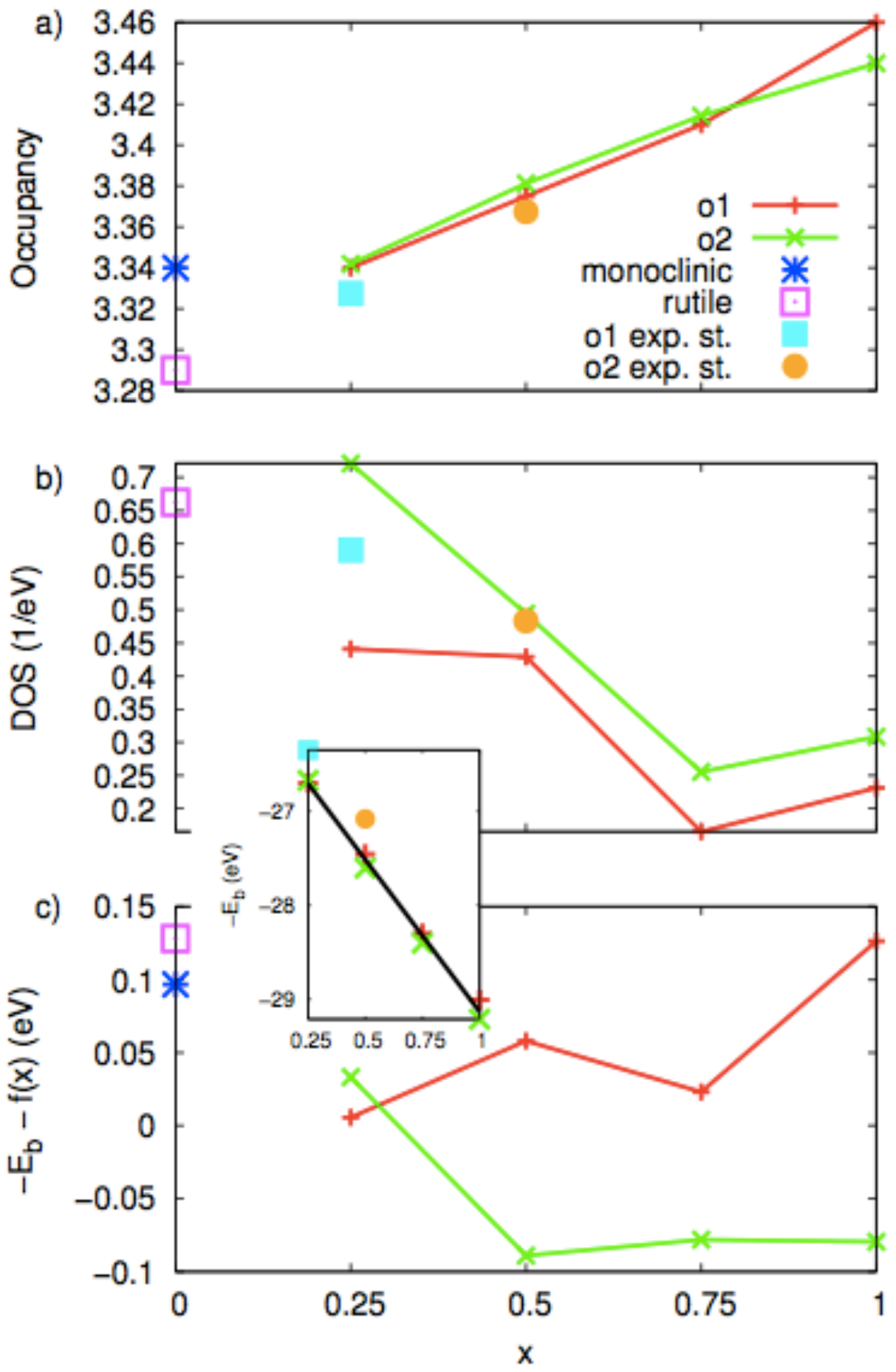

**Figure 9.** Results of *ab initio* calculations comparing the various structural phases at different hydrogen concentrations. a) Average occupancy of the vanadium 3*d* orbitals as calculated by Mulliken population analysis, plotted as a function of hydrogen concentration *x*. Values calculated from the experimental structure without geometry optimization are also included. b) Density of states at the Fermi level, calculated with a smearing width of 0.1 eV. Inset: Minus binding energy $E_b$ per $VO_2$ formula unit, in eV, plotted as a function of hydrogen concentration *x*. Linear fit, $f(x) = -3.235x - 25.898$, superimposed in black. c) Main panel: Binding energy with the linear trend removed to show the differences between phases. Note that the O1 phase is favored at $x = 0.25$, and the O2 phase is favored at higher values of *x*.

In order to address the issue of metallicity, it is instructive to examine the occupation of vanadium *d*-levels. As Figure 9a shows, the average *d*-level occupancy grows with hydrogen content, as discussed in prior work[4]. This can be easily understood as follows: the binding of the hydrogen to the oxygen shifts the electron density on oxygen away from the V-O-V bonds, as illustrated schematically in **Fig. 10**. Effectively, this is equivalent to the oxygen on site "B" being less electronegative, and as a result, the V-O bond becoming less ionic, with more electron density residing on the B-site vanadium compared to that in the rutile structure (site A). The de-tailed calculation of the electron occupancy using the Mulliken analysis corroborates this analysis quantitatively. Namely, we find the B-site oxygen to have larger Mulliken charge (6.68) compared with the A-site oxygen (6.57). This difference in charge $2\delta = 0.11$ is reflected in the two neighboring B-site vanadium atoms acquiring an additional charge $\delta \approx 0.06$ each, compared to the Mulliken occupancy of the A-site vanadium away from the O-H bond.

The less ionic $V_B..O_B$ bond results in a longer $V_B..O_B$ distance, compared to the $V_A..O_A$ bond away from the hydrogen site. This in turn leads to the $V_B..V_B$ distances being longer than the $V_A..V_A$ distances (see **Fig. 10**), explaining our experimental finding that the V..V distances are longer within the $V_2(OH)_2$ block of the O2 structure. We see that the periodic ordering of the H sites in the O2 structure thus leads to dimerization of the V..V distances, resulting in the doubling of the unit cell along the rutile *c*-axis. However, as mentioned earlier, the differentiation of V..V distances is less pronounced in the O2 phase than in the monoclinic phase of pristine $VO_2$, likely because of the overall expansion of the unit cell due to hydrogen absorption. We note that while theoretically, the O1 phase is also dimerized, in an experiment the disorder on hydrogen site results in only one inequivalent O position and the V..V dimerization is thus quenched.

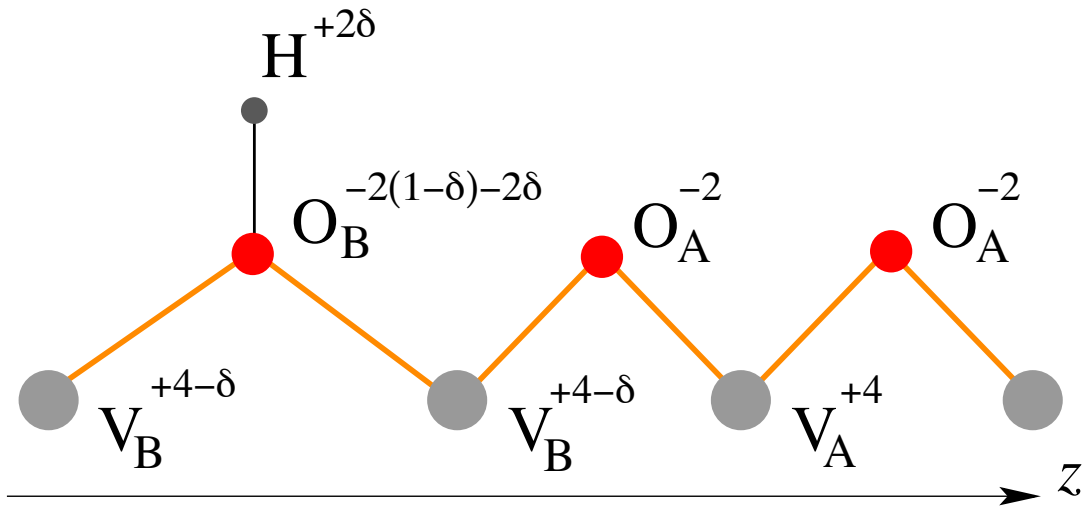

**Figure 10.** Schematic of charge redistribution due to hydrogen binding to B-site oxygens. This also results in the weaker $V_B..O_B$ bonds and consequently, in dimerization, with the $V_B..V_B$ distances being longer than the $V_A..V_A$ distances.

The increase in the *d*-level occupancy on B-site vanadium is reflected in the partial density of states (PDOS), so that the spectral weight shifts below the Fermi level, developing a pronounced peak, see **Fig. 11**a. By contrast, the A-site vanadium has PDOS reminiscent of the rutile structure[4], with a shoulder rather than a peak below the Fermi level. It is this increase in vanadium *d*-level occupancy that is responsible for the suppression of the metal-insulator transition in hydrogenated $VO_2$. Theory unambiguously predicts that both the O1 and O2 structures are metallic. Note however that despite the shift of the spectral weight below the Fermi level on B-site vanadium, the total DOS at the Fermi level actually decreases with increasing H doping (see Figure 9b). This is explained in part by the lattice expansion upon hydrogenation, which ought to decrease the DOS. The relative shifting of the vanadium *d* and oxygen 2*p* bands is also responsible for this trend, and the **Fig. 11**b shows the comparative PDOS due to the *p*-orbitals of oxygen on the A- and B-sites.

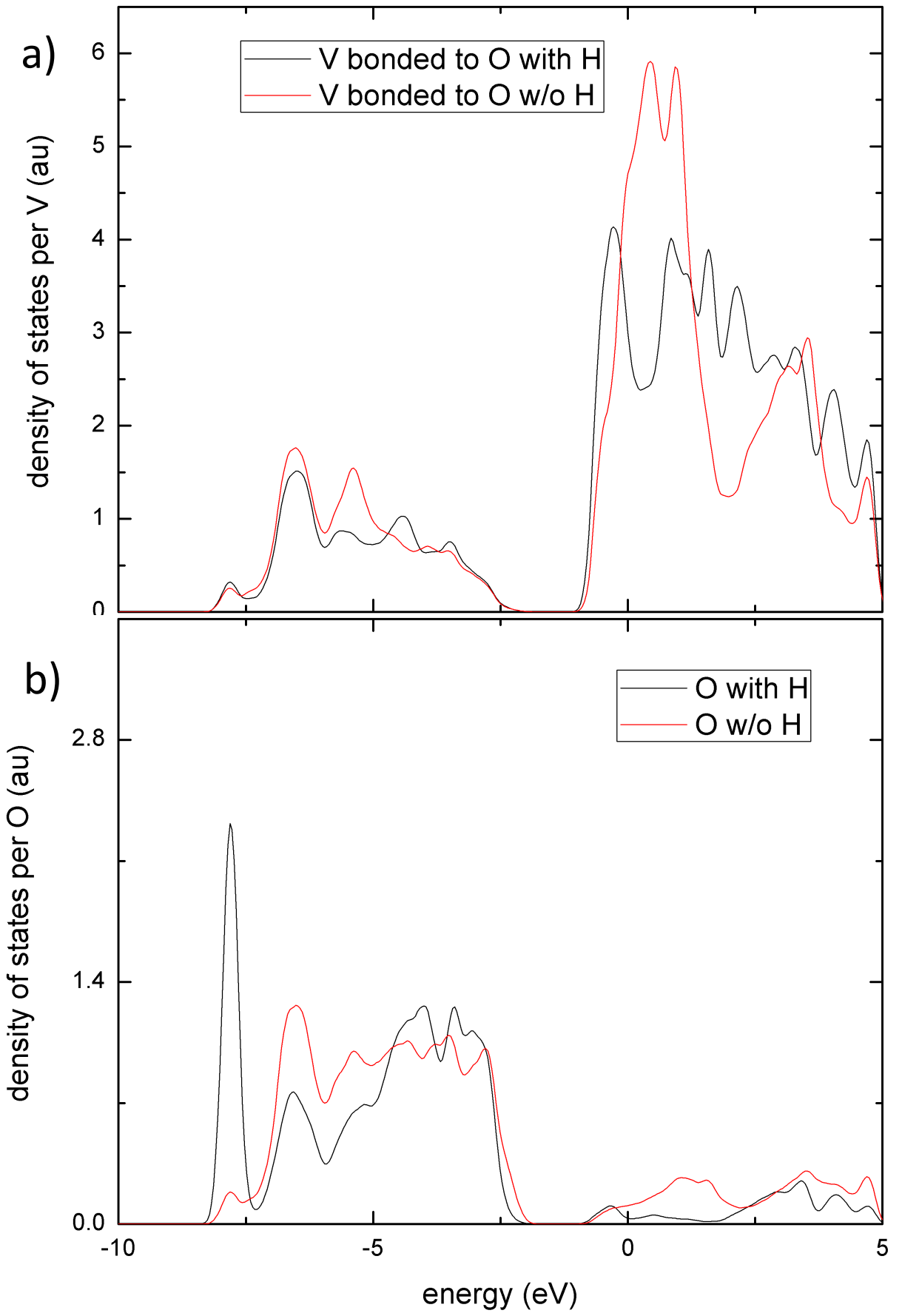

**Figure 11**. Calculated partial densities of states for (a) vanadium (3d band) and (b) oxygen (2p band) atoms. A-site atoms (no hydrogen proximal to the relevant oxygen) results are in red; B-site atoms (hydrogen proximal to the relevant oxygen) are in black. Note that hydrogen binding to oxygen results in a transfer of spectral weight below the Fermi level for the B-site vanadium 3d band.

The hydrogenation experiments give us insights into the nature and mechanism of the metal-insulator transition, which in pristine $VO_2$ is accompanied by dimerization of nearest-neighbor V..V distances in the low-temperature monoclinic phase. The results of the recent cluster-dynamical mean-field theory (CDMFT) calculations[17] on pure $VO_2$ corroborate earlier theoretical evidence[18] that the transition is not a purely Mott transition, but a result of the combination of the Peierls and Mott effects. Our findings indicate that the increase in vanadium *d*-level occupancy will suppress the tendency towards transition into the insulating Mott state. At the same time, the experimental results indicate that the O2 phase has V..V dimerization along the rutile *c*-axis. Nevertheless, theoretically both O1 and O2 phases remain metallic (confirmed in nanobeam measurements[4], though those experiments could not discriminate between O1 and O2), despite the dimerization, indicating that the (partial) depletion of the electronic density of states at the Fermi level due to the Peierls mechanism is not sufficient to stabilize the insulating phase when the vanadium *d* occupancy is moved away from half-filling. This can be explained in part by the expansion of the *c*-axis lattice parameter upon hydrogenation, which will lead to a decrease in the interatomic force constants and consequently, will reduce the energy savings due to dimerization that favors the Peierls state.

## CONCLUSIONS

In this work we report the stable structures adopted by vanadium dioxide as it accommodates atomic hydrogen as an intercalant. These phases were found through extensive variable temperature synchrotron powder diffraction experiments that examine the evolution of the $VO_2$ crystal structure *in situ* as it accommodates intercalated atomic hydrogen (supplied via catalytic spillover). We also report neutron powder diffraction of $D_xVO_2$/Pd material. None of these experiments show evidence of oxygen removal from the lattice. In addition to the previously identified structural phases (the M1 monoclinic $P2_1/c$ phase and the T tetragonal $P4_2/mnm$ phase, both of which can incorporate small amounts of hydrogen, as reported previously[3]; and the orthorhombic *Pnnm* phase of $H_xVO_2$ put forward based on inelastic neutron scattering experiments[2a]), we identify a new orthorhombic (O2) phase of *Fdd*2 symmetry (with a 2*a*,2*b*,2*c* supercell with respect to the *Pnnm* structure). Using orthorhombicity as a proxy for hydrogen concentration, calibrated via the neutron measurements of $D_{0.460(8)}VO_2$, we construct the phase diagram for the hydrogen/$VO_2$ system, as shown in Figure 8.

In addition to discovering the O2 phase, these experiments and their theoretical analysis give insight into the physics of the long-discussed metal-insulator transition in $VO_2$. First principles calculations confirm that one key effect of hydrogen doping is to increase the *d* occupancy of the vanadium ions bound to oxygens that are coupled to hydrogen atoms. This transfers spectral weight below the Fermi level in the electronic density of states, and thus favors metallicity. It is worth noting that the O2 phase, with its doubled unit cell, has a certain amount of dimerization of the one-dimensional vanadium chains that run along the *c* axis of the T and O1 phases. However, despite this dimerization, the O2 phase remains metallic, showing that Peierls physics and the resulting dimerization are not the dominant factors in determining the electronic structure of this phase. First principles calculations further confirm that at high *x* the O2 phase is expected to be energetically favored over the O1 phase.

These studies confirm that the hydrogen/vanadium dioxide system is rich, both from the point of view of interesting materials science and in terms of how doping affects the competition between phases in a strongly correlated material. We suggest several additional experiments, in light of these results and the revelation of the stability of the O1 and O2 phases. High resolution x-ray absorption fine structure (EXAFS) measurements could quantitatively confirm the theoretically inferred consequences of hydrogen's presence for the oxygen and vanadium effective valences. The resulting metallic O1 and O2 states are also targets for further study, particularly since the metallic state in pure $VO_2$ cannot be stabilized down to cryogenic temperatures. Given indications that the metallic T state in pure $VO_2$ is moderately correlated[19], the metallic $H_xVO_2$ states should be examined further through thermodynamic and transport measurements, to determine the properties of the charge carriers and the conventional/unconventional nature of the metallic states. Finally, more sophisticated theoretical techniques should be brought to bear on these phases, which present tests for approaches such as hybrid functionals[16]. Such correlated materials and a rich phase diagram are excellent proving grounds for advanced quantum chemistry approaches.

## ASSOCIATED CONTENT

Rietveld plots, unit cell and atomic parametes of described crystal structures, details of *ab initio* calculation are given in supplementary material. This material is available free of charge via the Internet at http://pubs.acs.org.

## AUTHOR INFORMATION

### Corresponding Authors

* yaroslav.filinchuk@uclouvain.be, nevidomskyy@rice.edu, natelson@rice.edu,

### Notes

The authors declare no competing financial interest.

## ACKNOWLEDGMENT

The authors thank the UCL and FNRS (CC 1.5169.12, PDR T016913) for financial support. We acknowledge the Fonds Spéciaux de Recherche (UCL) for the incoming postdoctoral fellowship co-funded by the Marie Curie actions of the European Commission granted to N. Tumanov. We thank ESRF and PSI for the beamtime allocation at the SNBL and MS beamlines and at the HRPT instrument. This project has received funding from the European Union's Seventh Framework Programme for research, technological development and demonstration under the NMI3-II Grant number 283883. DN, HJ, and JW acknowledge financial support of DOE BES award DE-FG02-06ER46337, and E. Morosan and Chih-Wei Chen for useful conversations and magnetization measurements.

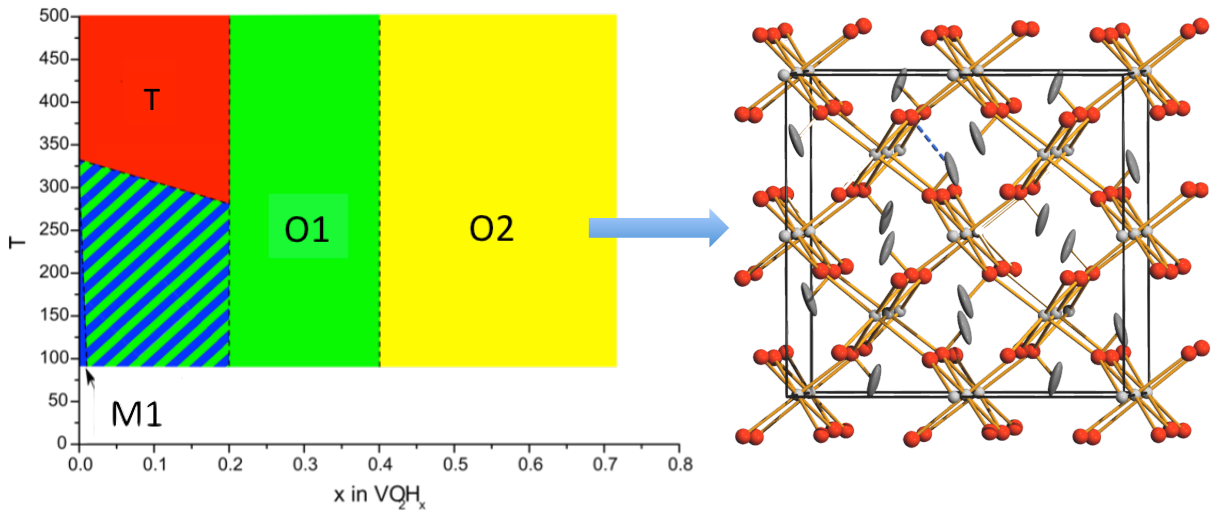

TOC figure

# *In situ* diffraction study of catalytic hydrogenation of $VO_2$: Stable phases and origins of metallicity

Yaroslav Filinchuk*, Nikolay A. Tumanov, Voraksmy Ban, Heng Ji, Jiang Wei, Michael W. Swift, Andriy H. Nevidomskyy*, and Douglas Natelson*

## Supporting Information

### 1. Detailed diffraction analysis

**Figure S1**. Rietveld refinement profile for the monoclinic phase at 80 K (related to Fig. 1 of the main manuscript)

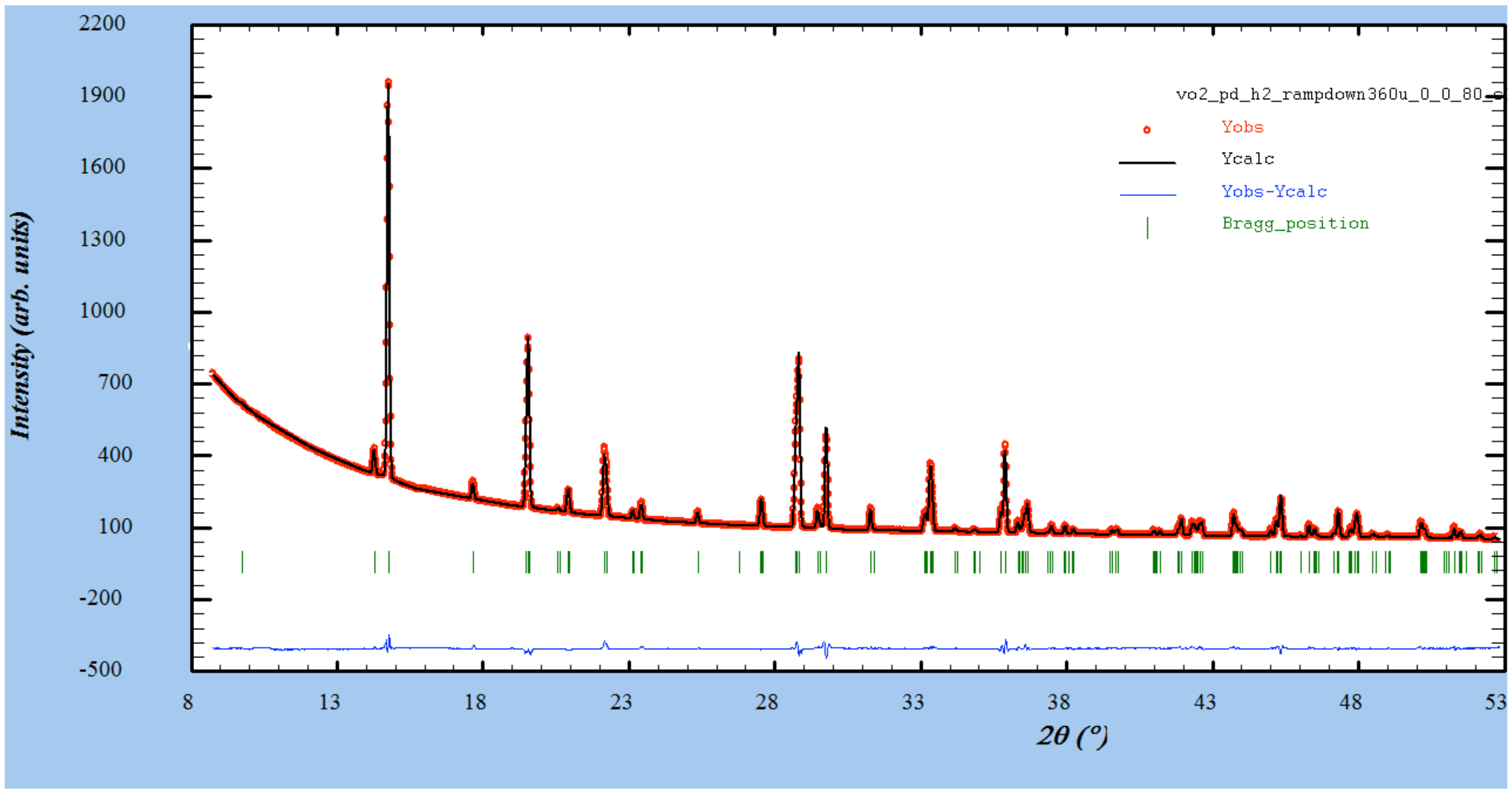

The refined parameters allow to quantify accurately the interatomic distances:

| Name | x | sx | y | sy | z | sz | B | sB | occ. | socc. | Mult |
|---|---|---|---|---|---|---|---|---|---|---|---|
| V | 0.26044( | 16) | 0.02198( | 13) | 0.28711( | 9) | 0.669( | 10) | 4.000( | 0) | 4 |
| O1 | 0.09462( | 68) | 0.29442( | 69) | 0.39682( | 54) | 0.868( | 45) | 4.000( | 0) | 4 |
| O2 | 0.60712( | 71) | 0.21003( | 69) | 0.39648( | 53) | 0.627( | 44) | 4.000( | 0) | 4 |

```
=> Cell parameters      :
                              5.35234    0.00006
                              4.52458    0.00005
                              5.37725    0.00006
                             90.00000    0.00000
```

```
                              115.33720   0.00074
                               90.00000   0.00000

Bragg R-factor:  4.32       Vol:  117.695( 0.002)

Coordination polyhedron for V atom:
(  V  )-( O1  ): 1.762(4)
(  V  )-( O1  ): 2.010(3)
(  V  )-( O1  ): 2.071(3)
(  V  )-( O2  ): 1.892(4)
(  V  )-( O2  ): 2.017(4)
(  V  )-( O2  ): 1.862(3)

(  V  )-(  V  ): 2.6139(8)
```

**Figure S2**. The Rietveld fit and structural data for the tetragonal $H_xVO_2$ $P4_2/mnm$ phase (data from Fig. 1 of the main manuscript for the sample held at 468 K).

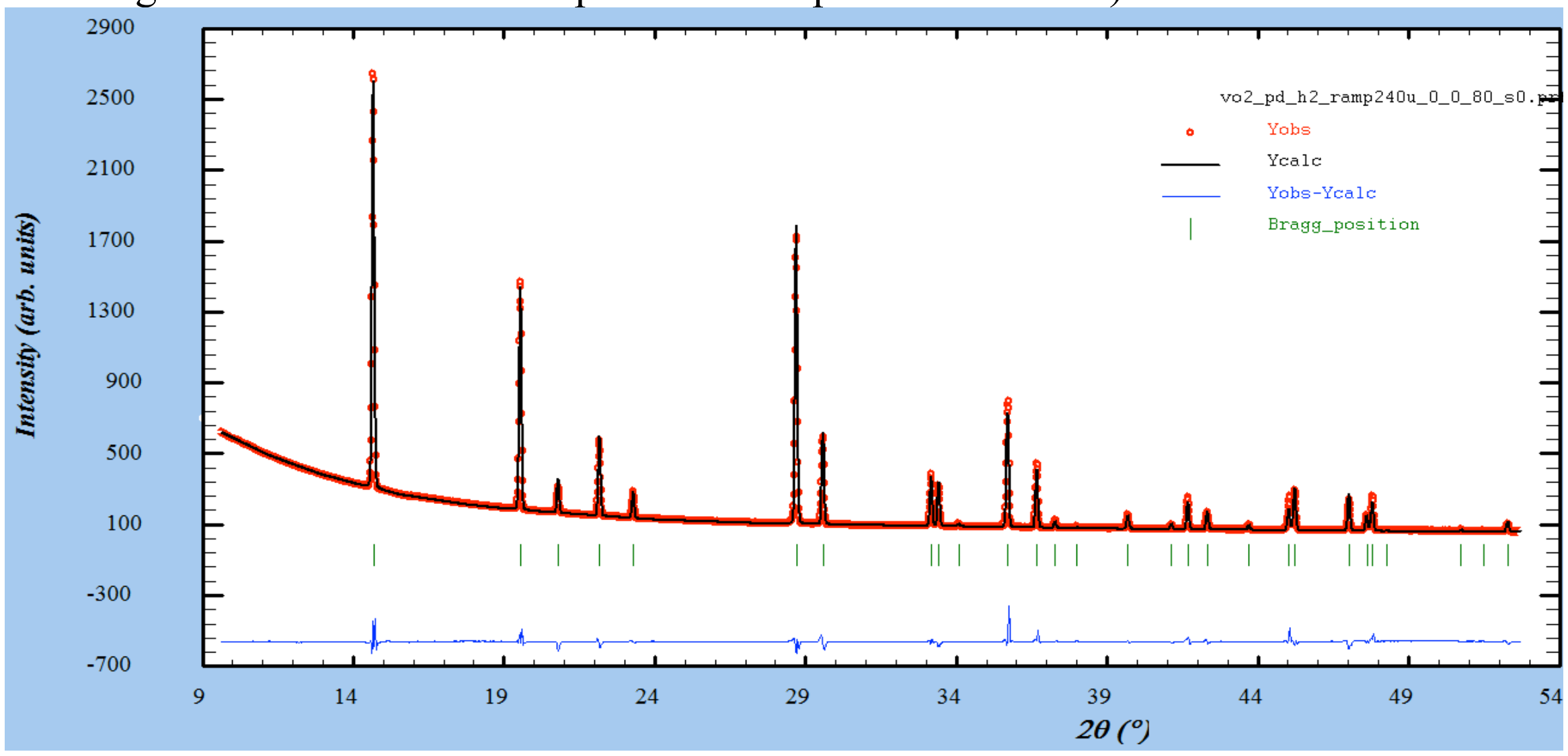

| Name | x | sx | y | sy | z | sz | B | sB | occ. | socc. | Mult |
|---|---|---|---|---|---|---|---|---|---|---|---|
| V | 0.00000( | 0) | 0.00000( | 0) | 0.00000( | 0) | 1.296( | 20) | 2.000( | 0) | 2 |
| O | 0.30207( | 29) | 0.30207( | 29) | 0.00000( | 0) | 1.102( | 42) | 4.000( | 0) | 4 |

```
=> Cell parameters      :
                            4.55659    0.00006
                            4.55659    0.00006
                            2.86076    0.00004

Bragg R-factor:  5.66       Vol:   59.397( 0.001)
```

Coordination polyhedron for Vatom is more regular than in the monoclinic structure:

```
(  V  )-(  O  ): 1.9467(14)
(  V  )-(  O  ): 1.9163(9)
(  V  )-(  O  ): 1.9163(9)
(  V  )-(  O  ): 1.9467(14)
(  V  )-(  O  ): 1.9163(9)
(  V  )-(  O  ): 1.9163(9)
```

**Figure S3**. Rietveld fit and refinement for the mixture of M1 and O1 phases initially present at 300 K in the as-prepared $H_xVO_2$ sample of Fig. 1 in the main manuscript.

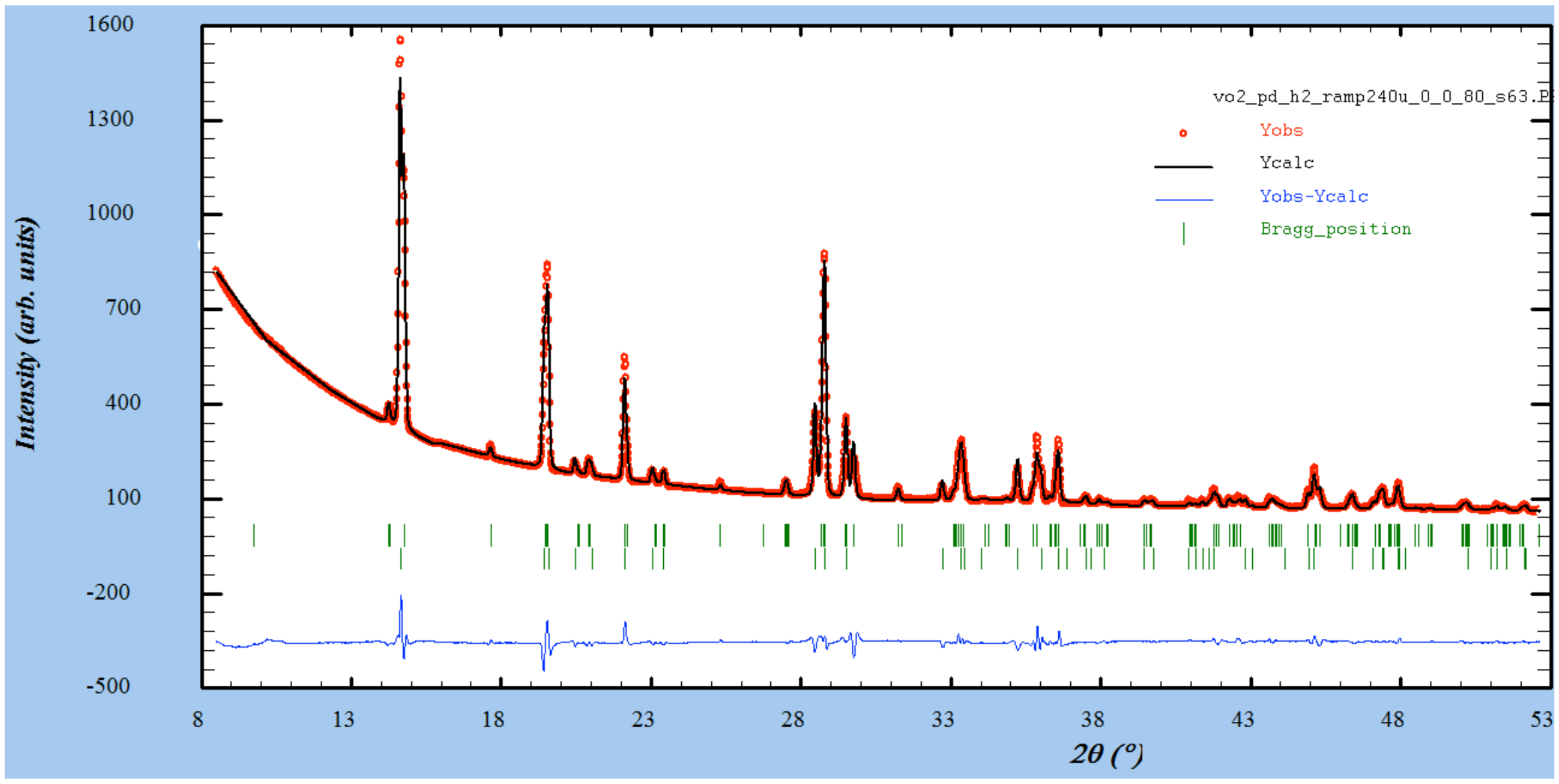

```
=> Phase No.  2 2 VO2Hx orth                           P n n m
  Name       x      sx        y      sy        z       sz       B    sB   occ.   socc.
Mult
 V        0.00000(   0)  0.00000(   0)  0.00000(   0)  1.314( 45)  2.000(  0)    2
 O        0.27895(  90)  0.31086(  84)  0.00000(   0)  0.879( 89)  4.000(  0)    4

=> Cell parameters      :
                              4.50611   0.00013
                              4.62997   0.00012
                              2.86721   0.00008

=> Phase:  1
 => Bragg R-factor:  6.78       Vol:  117.955( 0.011)  Fract(%):   50.05( 1.07)
 =>Rf-factor= 6.94             ATZ:        5308.172   Brindley:  1.0000

 => Phase:  2
=> Bragg R-factor:  8.11        Vol:   59.819( 0.003)  Fract(%):   49.95( 1.02)
=>Rf-factor= 6.64              ATZ:       10563.974   Brindley:  1.0000
```

Refined unit cell parameters of the O2 phase (obtained from $H_xVO_2$/Pd autoclaved under 15 bar $H_2$ at 468 K for three days):

_cell_length_a 9.4285(19)
_cell_length_b 8.9309(13)
_cell_length_c 5.7652(6)
_refine_ls_R_factor_all 0.0558
_refine_ls_R_factor_gt 0.0483
_refine_ls_wR_factor_ref 0.1331
_refine_ls_wR_factor_gt 0.1269
_refine_ls_goodness_of_fit_ref 1.066
_refine_ls_restrained_S_all 1.064

| Atom | *x* | *y* | *z* | *U* | ADP type | Occupancy |
|---|---|---|---|---|---|---|
| V1 | 0 | 0 | 0.4327(3) | 0.0201(5) | Uani | 1 |
| V2 | 0 | 0 | -0.0845(3) | 0.0259(9) | Uani | 1 |
| O1 | -0.0870(3) | 0.1167(3) | 0.176(3) | 0.0168(8) | Uani | 1 |
| O2 | -0.0898(3) | 0.1174(3) | 0.675(4) | 0.0169(8) | Uani | 1 |
| H1 | 0.0000 | 0.2047 | 0.1733 | 0.07(13) | Uiso | 0.26(18) |

**Figure S4**. Phase fractions for the M1, T and O1 phases in the in-situ experiment illustrated in the Figure 4. The experiment begins with approximately 50/50 mass fraction between O1 and monoclinic at 300 K (center of the plot), and the temperature cycling is as described in the main text and Fig.4.

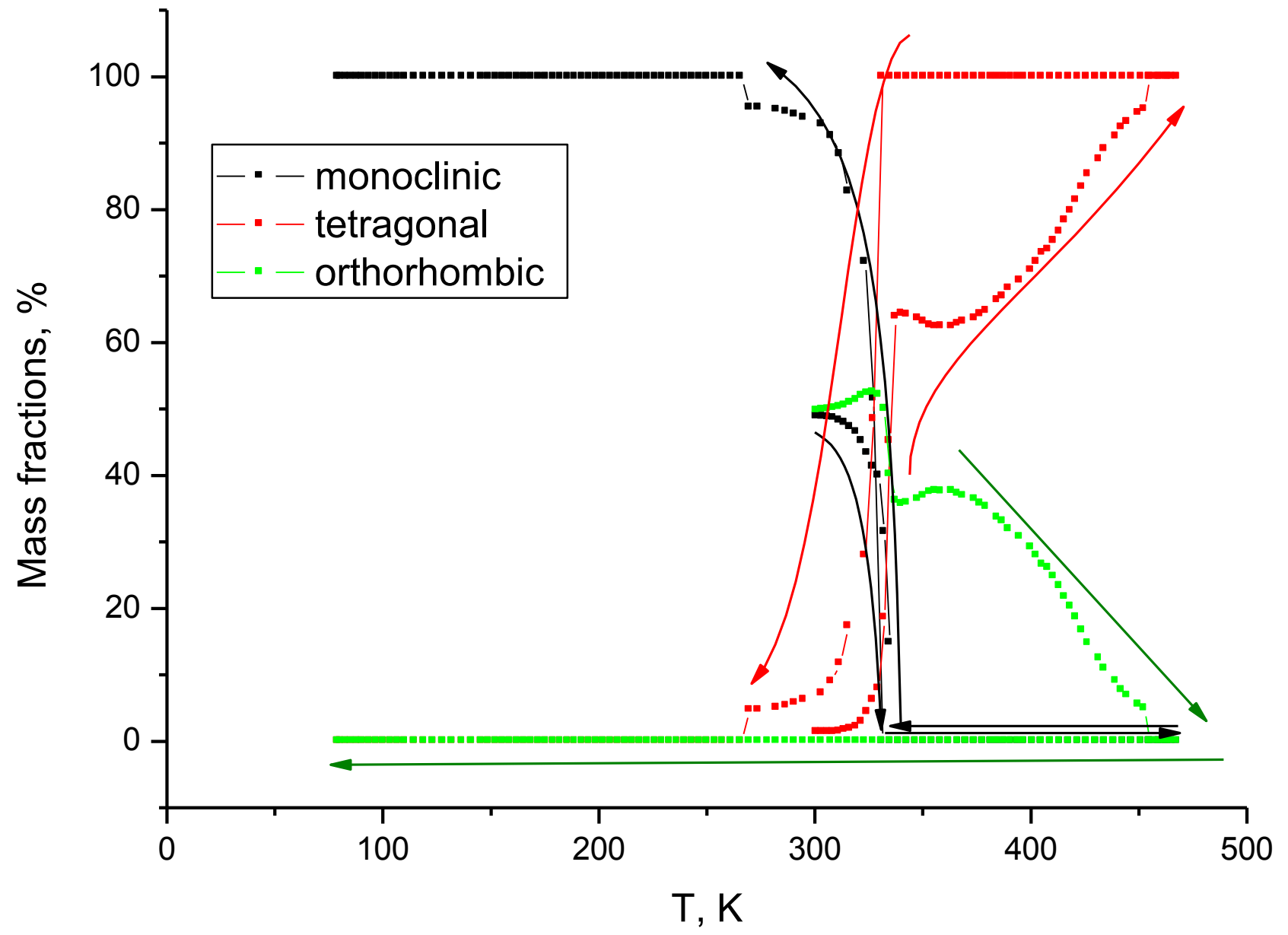

**Figure S5**. Rietveld fit for the O2 phase, modelled by two phases with identical structure and slightly different cell parameters. A zoom on the low and high angle parts is shown below.

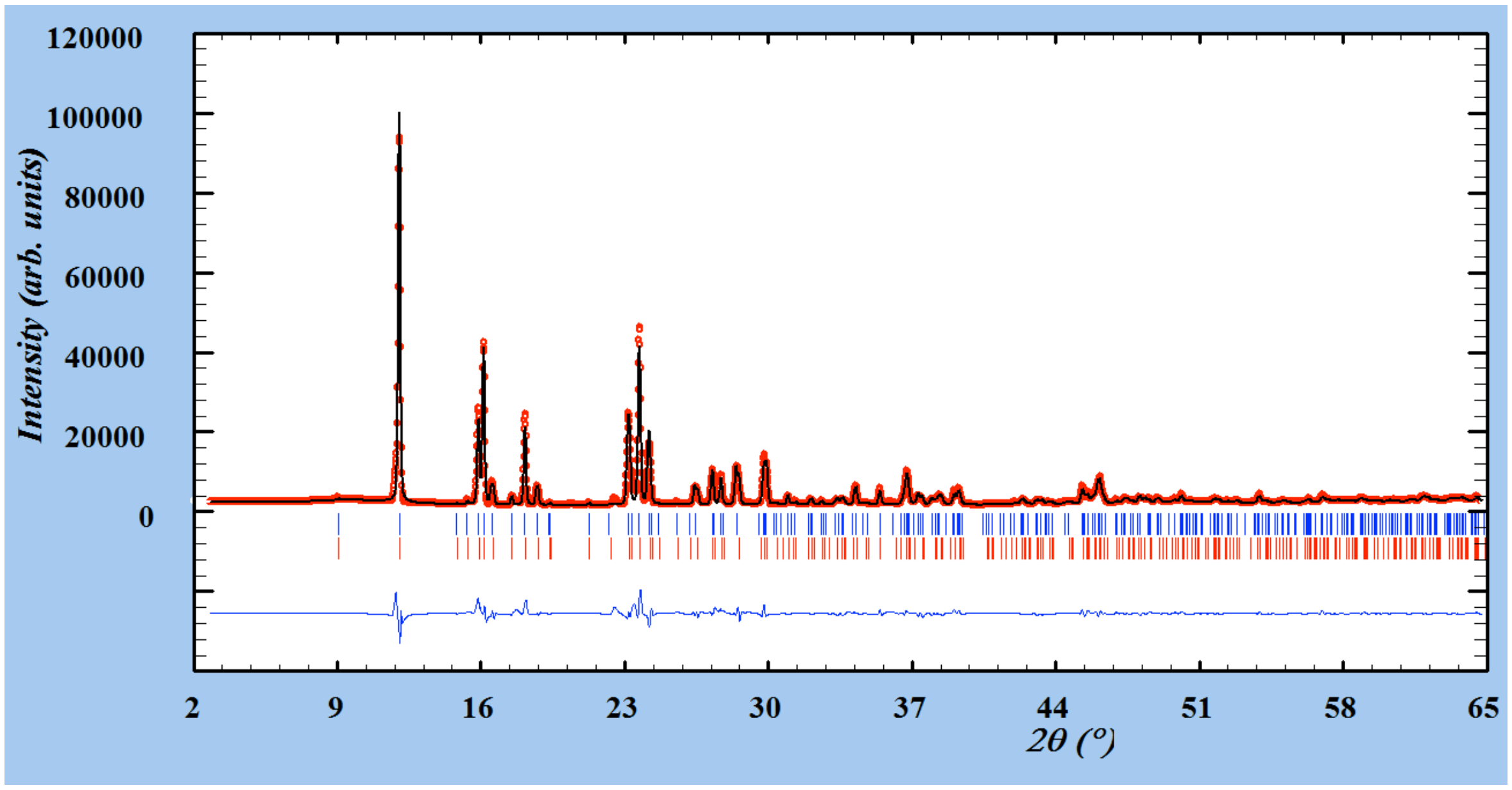

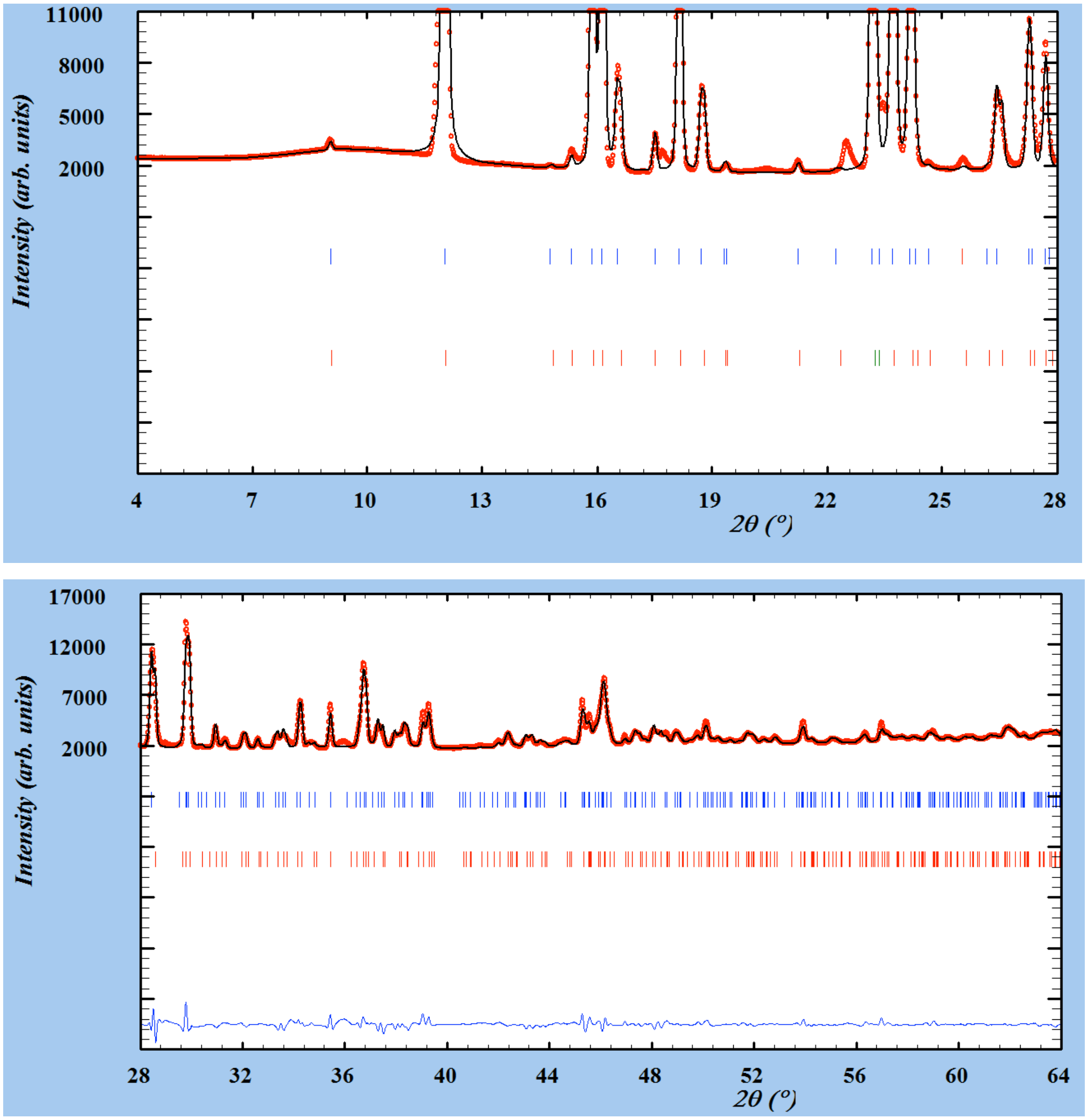

The final coordinates and quality of the fit from synchrotron X-ray powder data:

```
V1   0.00000   0.00000   0.43274   0.0138(3)   1.00000 Uiso V
 V2   0.00000   0.00000 -0.0846(8)   0.0138(3)   1.00000 Uiso V
O1  -0.0775(6)   0.1214(8)   0.1586(20)   0.0032(7)   1.00000 Uiso O
 O2  -0.0999(6)   0.1144(8)   0.6594(18)   0.0032(7)   1.00000 Uiso O
H -0.00170   0.20470   0.17330   0.02533   0.25000 Uiso H

 => Phase:  1      1 VO2Hx orth 2ax2bx2c
 => Bragg R-factor:   8.13       Vol:   493.566( 0.056)   Fract(%):   60.59( 1.30)
=>Rf-factor=  5.27             ATZ:       10616.346   Brindley:  1.0000
```

```
 => Phase:  2     2 VO2Hx orth 2ax2bx2c
=> Bragg R-factor:   9.37       Vol:  489.537( 0.071)  Fract(%):   39.41( 1.11)
=>Rf-factor=  5.76             ATZ:       10616.346   Brindley:  1.0000
```

The *z*-coordinate of the V1 atom was fixed in order to define the origin of the polar space group, and the H-atom was included into refinement from single crystal model, with all parameters fixed. Slightly different cell parameters for the two single crystals closely correspond to the slightly different cell parameters refined for the two O2 phases from powder data. The bimodal distribution suggests an incomplete homogeneity of the sample, which should improve on annealing.

The final coordinates and quality of the fit from neutron powder diffraction data:

V1  0.00000  0.43270  0.00000  0.01267  1.00000 Uiso V

V2  0.00000 -0.08450  0.00000  0.01267  1.00000 Uiso V

O1 -0.0889(4)  0.17600  0.1212(3)  0.0210(10)  1.00000 Uiso O

O2 -0.0883(5)  0.6811(13)  0.1120(3)  0.0233(11)  1.00000 Uiso O

D1 -0.0197(7)  0.144(2)  0.2060(9)  0.052(5)  0.461(9) Uani D

```
    _atom_site_aniso_U_11
    _atom_site_aniso_U_22
    _atom_site_aniso_U_33
    _atom_site_aniso_U_12
    _atom_site_aniso_U_13
    _atom_site_aniso_U_23
_atom_site_aniso_type_symbol
D1  0.019(4)  0.040(6)  0.097(6) -0.008(3) -0.030(4)  0.025(4)  D
_cell_length_a                         9.4610(6)
_cell_length_b                         5.7813(2)
_cell_length_c                         8.9705(3)
_cell_volume                           490.66(4)
```

=> Bragg R-factor:   8.30       Vol:  490.663( 0.040)  Fract(%):  100.00( 1.28)

=> Rf-factor=  5.45            ATZ:       1343.107   Brindley:  1.0000

**Figure S6**. Final Rietveld plot of the deuterated O2 phase sample

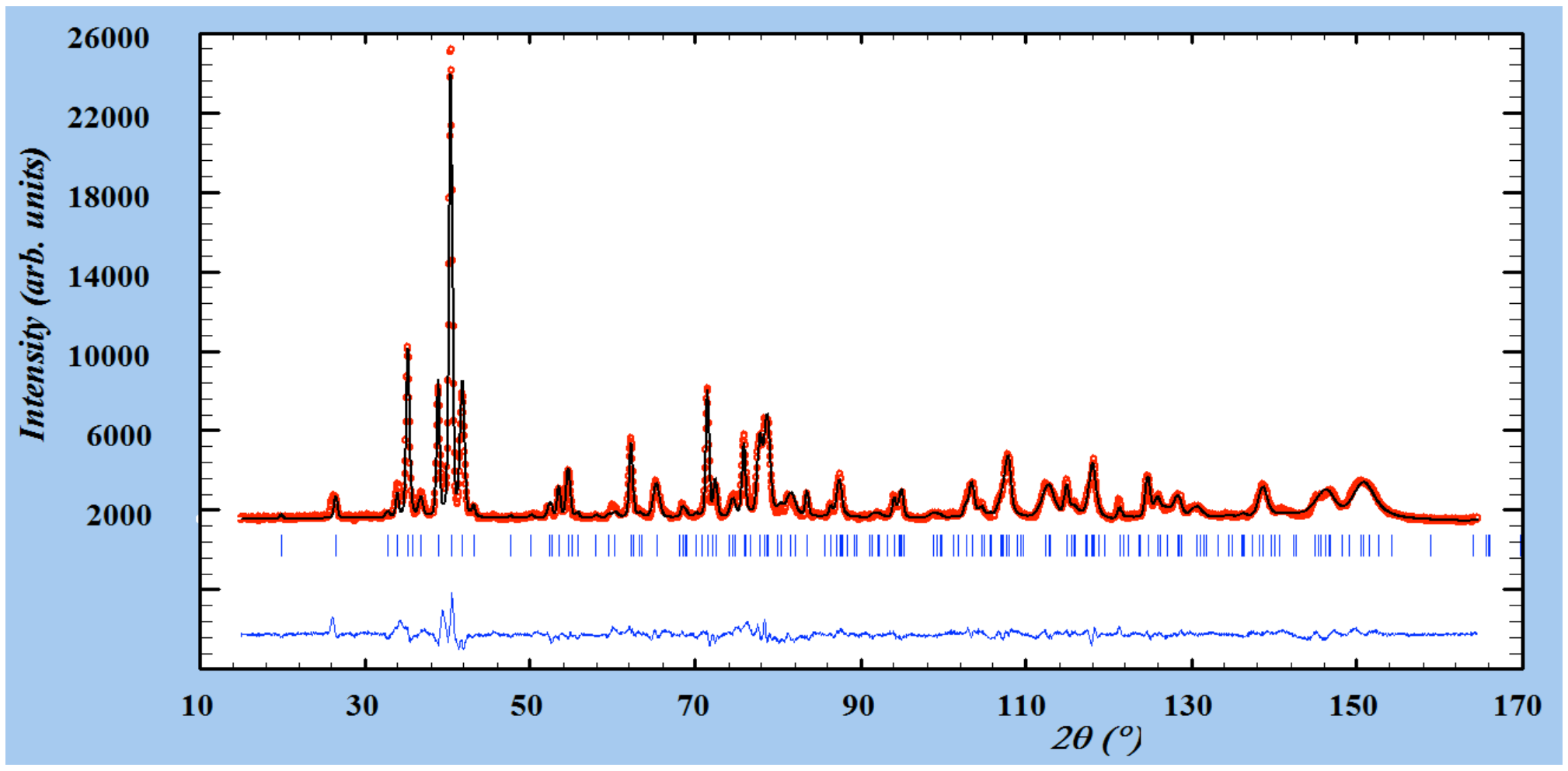

## 2. Ab initio calculations

Magnetic properties of the theoretical models were studied by performing spin-polarized geometry optimization calculations on structures set up with antiferromagnetic ordering wavevector Q = (0, 0, π). The converged structure had a higher binding energy, so the predicted ground state is antiferromagnetic. The AFM structures have higher binding energy than the PM structures by 50-200 meV per $H_xVO_2$ in both the O1 and O2 phases, lower occupancy by about 0.05 electrons and similar DOS per spin at the Fermi Level. However, since antiferromagnetic ordering has not been observed experimentally, we surmise that this is a spurious result similar to the unphysical AFM ground state predicted by GGA for pure $VO_2$ in the monoclinic phase[15].

**Figure S7**. Densities of states contributions for the tetragonal (left), O1 (center), and O2 (right) phases, as computed via the methods described in the main text. The Fermi level, found by overall charge neutrality, is set as the zero of the energy axis.

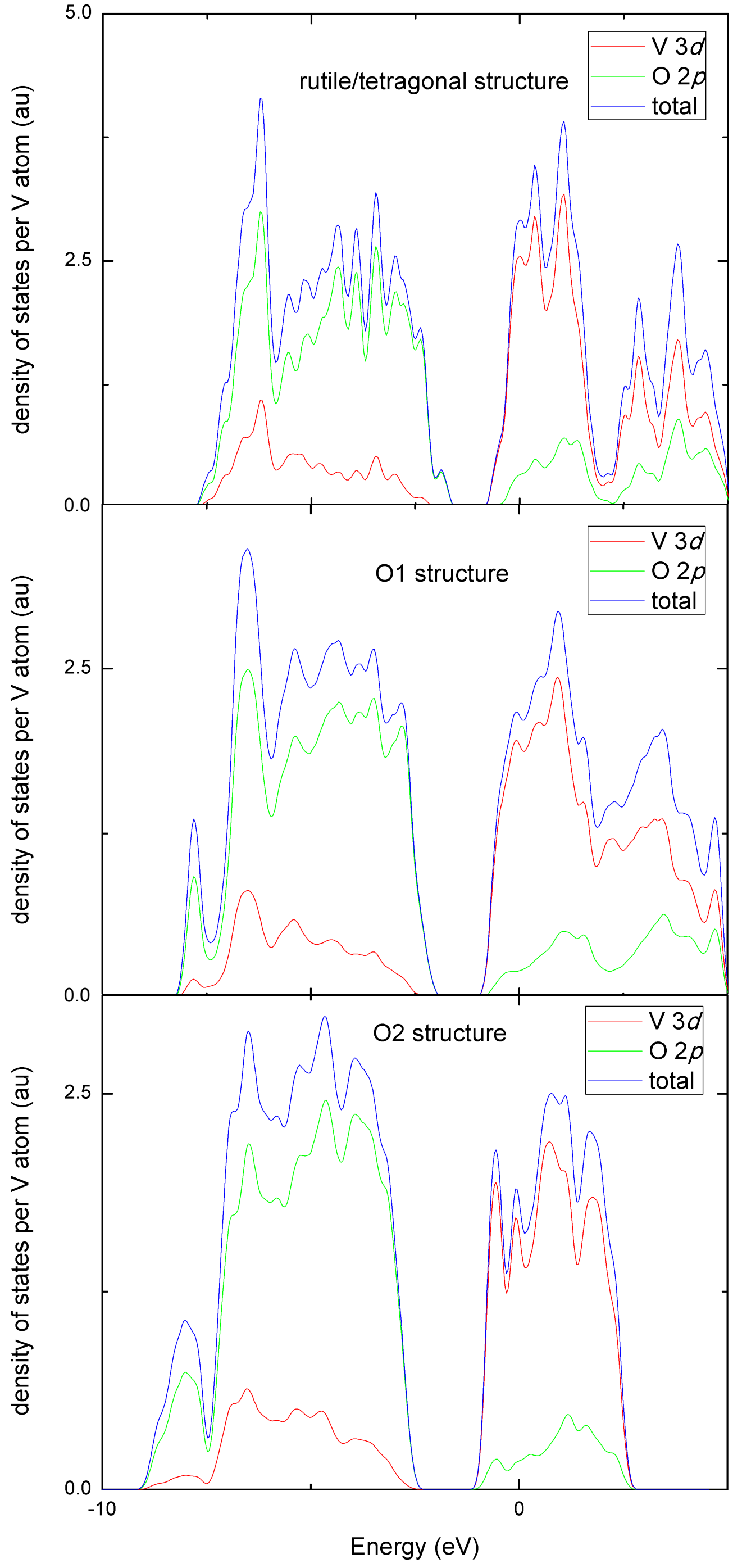